# Spontaneous Patterning of Binary Ligand Mixtures on CdSe Nanocrystals: from Random to Janus Packing

*Orian Elimelech,[a] Meirav Oded,[a] Daniel Harries,[a,b*] and Uri Banin[a*]*

a The Institute of Chemistry and The Center for Nanoscience and Nanotechnology, The Hebrew University of Jerusalem, Jerusalem 9190401, Israel.

b The Fritz Haber Center, The Hebrew University of Jerusalem, Jerusalem 9190401, Israel.

*E-mail: Uri.Banin@mail.huji.ac.il

Daniel.Harries@mail.huji.ac.il

ABSTRACT: Binary compositions of surface ligands are known to improve the colloidal stability and fluorescence quantum yield of nanocrystals (NCs), due to ligand-ligand interactions and surface organization. Herein, we follow the thermodynamics of a ligand exchange reaction of CdSe NCs with alkylthiols mixtures. The effects of ligand polarity and length difference on ligand packing were investigated using isothermal titration calorimetry (ITC). The thermodynamic signature of the formation of mixed ligand shells was observed. Correlating the experimental results with thermodynamic mixing models has allowed us to calculate the inter-chain interactions and to infer the final ligand shell configuration. Our findings demonstrate that the small dimensions of the NCs and subsequent increased interfacial region between dissimilar ligands, in contrast to macroscopic surfaces, allow the formation of a myriad of clustering patterns, controlled by the inter-ligand interactions. This work provides fundamental understanding of the parameters determining the ligand shell structure and should help guide smart surface design towards NC-based applications.

## MAIN TEXT

Surface ligands not only impact the chemical compatibility of nanocrystals (NCs)[1,2] but also provide passivation of surface dangling bonds,[3–5] thereby enabling control over the optical and electronic properties of the NCs.[6–8] Multi-component ligand coatings of NCs were demonstrated to improve their colloidal stability[9] and their fluorescence quantum yield.[10,11] Beyond its influence on the NCs' properties, the ability to engineer multi-ligand shell patterns provides important control over NCs self-assembly towards complex structures,[12] as well as their incorporation in many applications.[13]

The study of ligand patterning on inorganic NCs has mostly concentrated on Au nanoparticles (NPs),[14–17] and only few of these combined theory and experiment to resolve the interplay between inter-chain interactions and the resulting shell organization.[18,19] Surface ligand organization is dictated by the balance of configurational entropy, and the contribution of ligand-ligand interactions to the enthalpy and entropy of mixing.[15,18] By mixing chemically distinct ligands, it is possible to tune their surface organization, because full phase separation is typically enthalpically driven, while mixing between ligands is entropically favored.

Chemical distinction between mixed ligands can be introduced by altering chain polarity,[14–17,19] bulkiness,[20] and length.[9,18] For example, mixing short and long ligands liberates conformational and rotational degrees of freedom of the longer ligands unimpeded by the shorter chains at the interface.[9,18] This enthalpy-entropy balance leads to the final ligand patterning on the NC, which can show Janus, random mixing, or intermediate ligand clustering.[18] Unlike macroscopic systems, where phase separation is distinctly observed between regions of different composition, for microscopically small systems, the demixing transition is less sharp, mostly due to the relative increased contribution of the fraction of interfacial ligands residing in the region between ligand clusters. The ligand organization and the thermodynamics of these finite macroscopic systems is, therefore, more sensitive to details of the ligand-ligand interactions.[19]

To elucidate the inter-chain enthalpy-entropy balance at nanometric dimensions, we studied two main systems of NCs with a binary composition of capping ligands. First, we followed the effect of the ligands' length difference ($\Delta l$) on the inter-chain interactions, and as a result, on the distribution of the ligands within that shell. In the second system, we examined the effect of differences in degree of polarity between the ligands composing the shell, while maintaining similar ligand lengths.

In experiments, the binary shell was obtained via a ligand exchange reaction of oleate ($-O_2CR$) coated CdSe NCs with a mixture of linear alkylthiols (R'SH and R"SH) at different ratios (Scheme 1):

1. $$\mathrm{NC-O_2CR+(X)R'SH+(1-X)R''SH \rightleftarrows NC-(SR'_X SR''_{1-X})+RCO_2H}$$

As was reported for similar systems, the exchange reaction involves proton migration from the alkylthiol molecules to the oleate, thereby facilitating the latter's detachment from the $Cd^{2+}$ ion as oleic acid, and the attachment of the alkylthiolate, in accordance with the x-type ligand exchange mechanism. [21–24]

In the first system, designed to study the effect of variations in ligand length, Δl, several sets of binary shells were compared, each set consisting of 1-hexanethiol (C6SH), as the reference point, at varying ratios with respect to the longer ligands (i.e. 1-decaenthiol, C10SH; 1-tetradecanethiol, C14SH; 1-octadecanethiol, C18SH). Additionally, similar length differences were compared for different ligand lengths (i.e., C6SH:C10SH vs. C14SH:C18SH). In the second system, the C6SH ligand was mixed with 1H,1H,2H,2H-Perfluoro-1-hexanethiol (C6SH(F)). The thermodynamic parameters of the exchange reaction were extracted by Isothermal Titration Calorimetry (ITC) experiments and fitted to mixing models, allowing to resolve and dissect the different contributions to the thermodynamics of ligand shell patterning.

**Results and Discussion**

Figure 1 shows the ITC thermogram for the exchange reaction of oleate-coated CdSe NCs (d=3nm) at 303K with pure C6SH (a), pure C10SH (c), and a binary composition of $C6SH_{0.17}C10SH_{0.83}$ (b). All thermograms were normalized to the concentrations of $Cd^{2+}$ surface sites and the injected ligands. Integration of the thermogram peaks provides titration curves, representing the heat change as a function of the ratio between the added ligand and the $Cd^{2+}$ surface sites (Figure 1a-1c, overlay). All curves were fitted to a single-site exchange reaction model, presented in our previous work.[21] This model considers both the detachment of the native oleate ligand and the binding of the alkylthiol, similar to the approaches recently presented in other models for exchange reactions on the NC surface,[25–29] thus allowing to extract a single set of thermodynamic parameters, including the reaction enthalpy (ΔH), entropy (ΔS), and Gibbs free energy ($\Delta$G) (SI, section 4). Although the NC surface is expected to contain multiple types of surface sites,[30,31] this heterogeneity is apparently masked in the binary shell, probably due to the expected sparser packing (as will be discussed in the following) which resulted in thermodynamically identical facet and edge surface sites.

All measured reactions are exothermic ($\Delta \mathrm{H} < 0$) and favorable ($\Delta \mathrm{G} < 0$), as is expected for the exchange of a carboxylate binding group with a thiol. As reported elsewhere, the reaction involves an entropy loss ($\Delta S < 0$), mainly due to the replacement of the unorganized oleate ligand shell with the more organized alkylthiol ligand shell.[21]

As ligand chain length increases, the exchange with the pure ligand exhibits an increase in exothermicity (-20.2 and -24.3 kJ/mol for C6SH and C10SH, respectively) and entropy loss (-39 and -56 J/molK for C6SH and C10SH, respectively), leading to a minor change in $\Delta$G gain (-8.5 and -7.5 kJ/mol for C6SH and C10SH, respectively), in agreement with previous results.[21,22,27] For the exchange reaction with a binary mixture of ligands, the thermodynamic parameters are intermediate between the values for the two pure constituent ligands ($\Delta \mathrm{H} = -21.8\ \mathrm{kJ/mol}$, $\Delta S = -47\ \mathrm{J/molK}$ and $\Delta \mathrm{G} = -7.7\ \mathrm{kJ/mol}$). Closer inspection of these values reveals that although C10SH is the majority component, the ITC-extracted $\Delta$H and $\Delta S$ values more closely resemble those for C6SH, pointing to some contribution of the mixed ligand shell formation to the thermodynamics of the reaction.

To further unravel the effect of ligand composition on the extent of mixing, we analyzed the exchange with additional C6SH:C10SH ratios. Figures 1d, 1e, and 1f (green circles) present the extracted $\Delta$H, $\Delta S$, and $\Delta$G values of the measured compositions, respectively. Much like the exchange reaction with the pure ligands, the binary mixtures showed exothermic reaction heat and a loss of entropy. Moreover, as the molar fraction, $X_{C6SH,}$ of the C6SH increases, the exothermicity of the reaction decreases, along with the entropic loss, leading to overall only minor changes in ΔG. Similar enthalpy-entropy compensation (EEC) was previously observed in related NC systems (Figure S10).[21,22,32]

To resolve the contribution of ligand mixing, we compared the ITC-extracted results to a simple weighted linear combination (LC) of the pure ligands thermodynamic parameters (Figure 1d-f, green line). All calculations use the ligand ratio added to the solution. Due to the high and similar affinity (ΔG) of the pure alkylthiols to the NC surface, the final ligand ratio on the NC surface should be similar to the one in solution. Further justification for this assumption is presented in the SI, section 5 by using a Langmuir model for competitive adsorption.[33]

As the linear combination considers no mutual influence between both ligands, it represents the results expected for a fully phase separated system on the NC surface, whereas any deviation from this line is necessarily associated with the thermodynamic contribution due to mixing. This comparison reveals compensating positive deviations in the enthalpy (lower exothermicity) and in the entropy (lower entropy loss), resulting in overall negligible deviations in ΔG from the linear combination line.

Despite the similar polarity of both ligands, the mixing between the different lengths is enthalpically disfavored, yet entropically favorable, which can be ascribed to the increase in the number of microstates in the mixed system along the contact interfacial regimes. These trends are more pronounced with increasing molar ratio of C10SH (decreasing $X_{C6SH}$), suggesting higher perturbation of the C6SH ligands towards the C10SH ligand packing in the shell, than vice versa. From the perspective of a shell consisting of short ligands, embedded fractions of longer ligands should not significantly hamper inter-chain van der Waals (vdW) interactions. Indeed, the shorter ligand may even gain interactions with the longer ligand, due to possible folding-over of the longer ligands, which can lead to more compact packing, thus also lowering the entropy upon mixing. However, from the longer ligand's perspective, fractions of surrounding shorter ligands will reduce ligand-ligand vdW interactions, due to their smaller number of methylene groups, resulting in lower exothermicity ($\Delta H_{mix} > 0$). In addition, the presence of the shorter ligands provides additional rotational free volume to the longer chains, which directly affects their conformational entropy ($\Delta S_{mix} > 0$), as was observed by others.[9,18] Although the $\Delta G$ values we find are similar to the linear combination, suggesting no thermodynamic preference toward mixing ($\Delta G_{mix} \approx 0$), the observed deviations from linear combination, seen as lower exothermicity and entropy loss in the case of the C6SH:C10SH compositions, suggest some degree of mixing between both ligands. We note that while the solvent should contribute to the overall measured interactions in solution and on the NC surface, this contribution is already included in the calculated linear combination, thus the deviations in the parameters are attributed solely to the ligand-ligand interactions.

To further examine the interplay between mixing enthalpy and entropy, we studied a binary mixture of C6SH and C6SH(F). The largely different polarity between both ligands dictates a phase-separated ligand shell, as previously reported on Au NPs.[16,17] Figure 1g-1i presents the ITC-extracted thermodynamic parameters and the calculated linear combination. The pure C6SH(F) exhibits higher exothermicity and entropy loss relative to the hydrocarbon ligand, C6SH. The polar nature of the C-F bond induces stronger inter-chain interactions within the C6SH(F) ligand shell. In addition, it has been observed that fluor-rich ligands prevent the penetration of small molecules into the ligand shell, due to unfavorable interactions between the fluorocarbon and hydrocarbon molecules.[34–36] Both effects suggest compact ligand packing, with increasing vdW interactions, leading to elevated exothermicity and entropy loss.

For all the investigated C6SH:C6SH(F) compositions, the measured ΔH and $\Delta S$ values mostly correlate with the linear combination (Figure 1g and 1h, respectively), while the measured ΔG values present positive deviations from the linear combination (Figure 1i), suggesting overall unfavorable mixing. The limited contact interfacial region between clusters of each ligand results in small deviations from the linear combination for ΔH and $\Delta S$. These deviations demonstrate a slight reduction in the exothermicity and a small increase in the entropic loss, as expected for the unfavorable mixing at the interface. In contrast to the C6SH:C10SH mixture, all extracted thermodynamic parameters collectively point to an assembly pattern where, due to their different polarity, the ligands tend to fully segregate and form distinct domains of each ligand on the NC shell (Janus-like surface pattern, Scheme 1).

To gain further insight to the link between ligand-ligand interactions and the resulting shell organization, we analyzed the experimental findings using several mixing thermodynamic models. Because experiments show a non-zero mixing enthalpy, the simplest “ideal mixture” model can be excluded. Instead, the mixture can be treated using the "regular solution" model, where the mixing free energy, $\Delta G_{mix}$, can be expressed as:

$$2. \qquad \Delta G_{mix} = \Delta S_{mix}^{conf,id} + \xi_G X_1 X_2$$

In Eq.2, the first term on the right-hand side is the configurational entropy, $\Delta S_{mix}^{conf,id}$, associated with the organization of the ligands on the NC surface, and assumed to be ideal in this mean field model (SI, section 6). The second term is the nonideal contribution, with $\xi_G$ the free energy change involved in pairing two different ligands from pairs of similar ligands, which can have enthalpic and entropic components,

3. $$\xi_G = \xi_H - T\xi_S$$

Thus, $\xi_G$ embodies a nonideal mixing enthalpy $\Delta H_{mix}$ contribution and an additional non-configurational entropy, $\Delta S_{mix}^{nc}$, associated with the intra- and inter-chain degrees of freedom, as was previously described for mixtures of long and short ligands[9,18] (SI, section 6).

For the C6SH:C10SH system, a good fit was achieved with $\xi_H = 8 \pm 3$ kJ/mol (Figure 1d, blue line), resulting in a maximal value of $\Delta H_{mix} = 2 \pm 1$ kJ/mol, for $X_{C6SH}$=0.5. Positive enthalpic interaction parameter $\xi_H$ is consistent with a loss of interactions due to C6SH:C10SH pairing. The difference in measured enthalpy between the pure C6SH and C10SH ligands is 1 kJ/mol per carbon, thus the expected maximal loss of end-chain interactions for C10SH upon mixing with C6SH is 4 kJ/mol. Yet, we find a lower value for $\Delta H_{mix}$, indicating that either the C10SH ligands bend towards the NC surface,[21,37] so as to compensate for the loss of end-chain interactions, and/or that inhomogeneous mixing is realized with partial ligand segregation (i.e., ligand clustering), leading to lower loss of interactions because of the smaller interfacial regions.

Considering $\Delta S_{mix}^{conf,id}$ as the sole entropic term (i.e., setting $\xi_S = 0$), already provides a good fit to the ITC-extracted ΔS (Figure 1e, blue line), thus suggesting a well-mixed ligand shell. However, $\Delta G_{mix}$, derived as the sum of fitted $\Delta H_{mix}$ and $\Delta S_{mix}$, mostly results in positive deviations from the linear combination over most of the measured molar fraction range (Figure 1f, blue solid line). In the mean-field representation, ΔG for the ligand exchange reaction can then be determined by the common tangent construction (Figure 1f, blue dashed line). The region between the two contact points on the $\Delta G$ curve should correspond to meta- or unstable states that should evolve to phase separation between the two ligands. However, this result is inconsistent with the known similar polarity of both ligands and the observed gain in $\Delta S_{mix}$. Indeed, previous works on binary ligand shells of varying hydrocarbon lengths suggested some degree of ligand segregation, resulting in the formation of clusters on the NC surface (Scheme 1), but not full phase separation.[9] This apparent inconsistency may be a consequence of the small size of the NC surface (127 surface sites), since finite size effects are unaccounted for in simple mean-field models.

To account for the shortcomings of mean-field representation, we used a Monte-Carlo based simulation with thermodynamic integration[38] to numerically evaluate $\Delta G_{mix}$, $\Delta H_{mix}$ , and $\Delta S_{mix}$ for a finite-sized system (Figure 1f, red. SI, section 6). The model considers pairwise interactions $\chi_G$ between nearest neighbors on a square lattice. This interaction includes enthalpic and entropic components of the form:

4. $\chi_G = \chi_H - T\chi_S$

Fits to the experimental results yielded interaction parameters of: $\chi_G = 3.4RT$, $\chi_H = 15.4\ kJ/mol$, and $\chi_S = 23\ J/molK$, from which $\Delta H_{mix}$ and total $\Delta S_{mix}$ were calculated (Figure 1d and 1e, respectively, red), while also accounting for the exact $\Delta S_{mix}^{conf}$ (Figure 1e, black. SI, section 6). Both $\chi_H$ and $\chi_S$ are positive, which points to enthalpically unfavorable mixing compensated by the entropic tendency to mix. However, the corrected conformational mixing entropy is lower than the ideal one ($\Delta S_{mix}^{conf} < \Delta S_{mix}^{conf,id}$), suggesting partial ligand segregation and the formation of clusters. The same procedure applied to the C6SH:C6SH(F) binary set indicates a higher $\chi_G$ of 5.2RT (Figure 1i, red).

Figure 2 compares simulation results for mixtures of two different ligands, $L_A$ and $L_B$, with molar fraction X=0.5 and interaction parameter $\chi_G$ on the NC surface (11×11 =121 sites) with a macroscopic system (111×111, similar results were achieved for a larger 1001×1001 grid) of the same composition. The black line, representing the NC system, does not show phase transition, but rather a continuous and gradual progression between different ligand surface patterning, as shown with insets. Zero interaction parameter ($\chi_G = 0$) represents fully compensating enthalpy and entropy interaction parameters, so that the only driving force for mixing is the configurational entropy ($\Delta G_{mix} = -T\Delta S_{mix} = -T\Delta S_{mix}^{conf,id}$), and the ligands distribute randomly with 50% of $L_A$-$L_B$ pairs on average. Note that even random mixing (compensating $\Delta H_{mix}$ and $\Delta S_{mix}^{nc}$) should lead to some apparent clustering of ligands on the NC surface (Figure 2b). For the C6SH:C10SH mixture, we find $\chi_G = 3.4RT$, the enthalpic preference towards demixing overcomes the entropic preference towards mixing, and the resulting clusters are larger, with only 25% of $L_A$-$L_B$ pairs on average. This agrees with what has previously been observed for a binary shell with varying ligand lengths.[9] Thus, compensation between the reduced enthalpy (loss of inter-chain interactions) and increased entropy dictates the extent of clustering.

For the C6SH:C6SH(F) mixtures, $\chi_G = 5.2RT$, and mixing between the ligands is unfavorable. Here we find two well separated regions of ligands with only 15% of $L_A$-$L_B$ pairs on average associated with the limited interface area. The opposite limit of phase-separated ligand shell can potentially be achieved for negative values of $\chi_G$ leading to homogenous ("checkerboard pattern") organization driven by the favorable $L_A$-$L_B$ mixing, which we estimated at 90% $L_A$-$L_B$ pairs for $\chi_G = -6RT$.

The larger influence of the interfacial fraction for small, NC-like systems can be appreciated by comparing it to a macroscopic system approaching the thermodynamic limit. For the larger grid, a sharper change is observed between $\chi_{\mathrm{G}} = 2\mathrm{RT}$ and $\chi_{\mathrm{G}} = 4\mathrm{RT}$ (Figure 2a, grey stars), in agreement with the known critical point,[39,40] where the $L_A$-$L_B$ pair fraction is expected to decrease abruptly with increasing $\chi_{\mathrm{G}}$. Similarly, the average area of the simulated clusters swiftly changes around the critical point in the macroscopic system, while the simulated NC system presents only a moderate and gradual change (Figure 2b and Figure S13).

The compensation between $\chi_{\mathrm{H}}$ and $\chi_{S}$ was further studied for additional alkylthiol binary systems of a similar reference point, C6SH, with C14SH, ($\Delta l = 8$; Figure 3a-3b) and with C18SH ($\Delta l = 12$; Figure 3c-3d). For both compositions, the extracted values for ΔG show minor deviations from the linear combination (Figure S10), and a similar $\chi_{\mathrm{G}}$ to the one calculated for the C6SH:C10SH mixture ($\chi_{\mathrm{G}} = 3.4\mathrm{RT}$). Hence, ligand clustering is also expected in the C6SH:C14SH and C6SH:C18SH mixtures. Similar trends of positive deviations from linear combination in ΔH and ΔS were observed. Yet, these deviations become larger as Δl increases. The maximal deviation in ΔH increases from 1.8 through 3 to 5.5 kJ/mol as Δl grows from 4 through 8 to 12, respectively. Correspondingly, $\chi_{\mathrm{H}}$ is positive (endothermic mixing) and increases with increasing Δl, due to a higher loss of end-chain interactions at the interfaces between both ligands (Figure 3e). Likewise, the maximal deviation in ΔS increases from 6 through 9.5 to 20 J/molK as Δl increases from 4 through 8 to 12. We find that the fitted positive $\chi_{\mathrm{S}}$, compensating for the loss of interactions by increasing the entropic gain as Δl grows due to the additional release of degrees of freedom for the longer ligands at the interfacial contacts (Figure 3f).

Careful examination of the experimental results reveals slightly higher deviations from the linear combination and from the fit for low $X_{C6SH}$, as was observed for the C6SH:C10SH system above. This asymmetric behavior is even more pronounced for mixing two long ligands: C14SH and C18SH (Figure 4). Unlike the C6SH:C10SH system with similar $\Delta l = 4$, for which a monotonic behavior is observed upon changing $X_{C6SH,}$ for the C14SH: C18SH mixture, above a certain $X_{C14SH}$ the contribution of mixing becomes exothermic (Figure 4a) with higher entropy loss (Figure 4b). For low $X_{C14SH}$, the mixing enthalpy (or mixing entropy) is mainly affected by the loss of interactions (or increase in degrees of freedom) of the longer C18SH, which is the majority component. Therefore, mixing is endothermic with positive entropic contribution, as observed for the previously discussed systems. For high $X_{C14SH}$, however, the short chain, which dominates the mixing characteristic, gains much more interactions when mixing with C18SH, thus forming a tightly packed ligand shell and consequently losing entropy through mixing. This is also potentially related to the previously observed tendency of long ligands to lie flat on the NC surface.[37] The long C18SH better overlaps the C14SH chain as both of them tend to fold.

To explain these experimental results molar fraction-dependent interaction parameters are required. We define interaction parameters that depend linearly on the molar fraction of the components, similar to the ones suggested by the sub-regular solution model.[41,42]

5. (a) $\chi_H = \omega_{14,18}X_{14} + \omega_{18,14}X_{18}$ (b) $\chi_S = \eta_{14,18}X_{14} + \eta_{18,14}X_{18}$

where $\omega_{i,j}$ and $\eta_{i,j}$ represent the gain in enthalpy and non-configurational entropy, respectively, resulting from the insertion of ligand j to a region containing ligand i. Eq. 5 results in a better fit of the ITC-extracted $\Delta H$ and $\Delta S$ (Figure 4a and 4b, respectively) while $\Delta G_{mix} \approx 0$, as observed experimentally (Figure S10), allowing us to use the same $\chi_G = 3.4RT$. Interestingly, the coefficients $\omega_{14,18}$ and $\eta_{14,18}$ are negative (-108 kJ/mol and -117 J/molK, respectively) while the coefficients $\omega_{18,14}$ and $\eta_{18,14}$ are positive (86 kJ/mol and 78 J/molK, respectively, Figure 4c and 4d). The sign change supports our conjecture regarding the gain in interactions and loss of non-configurational entropy for the shorter C14SH when in proximity to the longer C18SH, and vice versa for C18SH.

**Conclusions:**

Using ITC, we were able to follow the thermodynamics of ligand exchange reaction from oleate-coated NCs towards a binary composition of alkylthiols. Our findings revealed that binary compositions with similar polarity exhibited a compensation mechanism between enthalpy and entropy of mixing, leading to almost no change in free energy upon mixing. Using simple models, we resolved the entropic and enthalpic contributions due to ligand-ligand interactions, as well as the contribution of the configurational entropy on the NC surface. Larger differences between the mixed ligand chains resulted in higher mixing endothermicity, compensated by an increase in the mixing entropy. This trend is mostly due to the loss of interactions at the ligand-ligand contact interfacial zone between different ligand domains, which increases with increasing ligands' length difference (Δl). For systems with similar Δl but longer ligand lengths, we find a non-monotonic change in the mixing enthalpy and entropy, which we rationalize by a difference in the thermodynamic response of both ligands to changes in their surroundings. Based on the calculated total interaction parameter $\chi_G$ values, we could infer the mixed shell structure, which is most strongly affected by the chemical nature of the ligands (hydrocarbon vs. fluorocarbon). Our findings demonstrate that the small dimensions of the NCs and subsequent increased interfacial region between non-similar ligands, in comparison to macroscopic surfaces, allow the formation of myriad of clustering patterns, controlled by the inter-ligand interactions. As a rule of thumb, we find that clustering is expected for a mixture of alkylthiol ligands with similar polarity but with different ligand length. By contrast, binding of mixed ligands with different polarities (hydrocarbon and fluorocarbon) but similar lengths, resulted in their separation into segregated clusters on the NC surface. Our findings aid in the fundamental understanding of the formation of surface ligand clustering versus phase separated Janus-like surface arrangement, which should allow smarter surface design towards NC-based applications.

**Experimental**:

**Chemicals**: 1-Octadecene (90%), Oleic acid (90%), CdO (≥99.99%), Se powder (100 mesh, 99.99%), Trichloroethylene (anhydrous, ≥99%), 1-Hexanethiol (97%), and 1-Octadecanethiol (98%), were purchased from Sigma Aldrich. 1-Decanethiol (96%) and 1-Tetradecanethiol (94%) were purchased from Alfa Aesar. 1H,1H,2H,2H-Perfluoro-1-hexanethiol (97%) was purchased from Synquest laboratories. Hydrochloric acid (37%) and HPLC grade hexane were purchased from Bio-Lab.

**Methods:** Absorption measurements were performed using JASCO V-770 UV-vis-NIR spectrophotometer. ITC experiments were performed using a NanoITC calorimeter (TA instruments) equipped with 1ml Hastelloy sample and reference cells with a 250 µl syringe. TGA measurements were performed using TGA-5500 (TA instruments). The ITC curves were fitted using NanoAnalyze Software v 3.10.0 (TA instrument). Further data analyses including mixing model fitting and ligand shell structure calculations were done using MATLAB (The MathWorks Inc.).

**CdSe NCs synthesis**: CdSe NCs (d=3nm) in a zinc blende structure were synthesized by modifying a known procedure.[10,22] Briefly, in a 100 ml three-neck flask 4 ml of 0.2 M Cd-oleate and 13 ml of ODE were degassed under vacuum at 100°C for 1 hour. Then the temperature was increased to 240°C under Ar flow and 4 ml of 0.1 M Se suspension in ODE was quickly injected. An additional 200 µl of 0.1 M Se-ODE suspension was injected every 5 minutes to avoid Ostwald ripening. The NCs reached the desired size (d=3 nm) after 30 minutes. The NCs were precipitated from the synthesis crude solution by centrifugation with toluene and ethanol at 60K RPM for 10 minutes. Then, the NCs were re-dispersed in toluene and precipitated again with ethanol. The last step was repeated three times in order to get rid of excess ligands. The NCs clean solution was kept in trichloroethylene (TCE) under Ar atmosphere. The NC size was determined from a previously reported sizing curve based on the first exciton peak position (Figure S1).[43,44]

**ITC measurements:** Alkylthiol ligands and purified NCs dispersed in TCE were used for ITC measurements. This solvent was chosen due to its relatively high boiling point, and a relatively low enthalpy of mixing with the ligands.[45] In addition, all investigated ligands and NCs are well dispersed in TCE. The NCs concentration was determined from the solution absorption, based on a previous report of the extinction coefficient.[43] The surface sites concentration was calculated based on a simple spherical model of zinc blende CdSe with lattice parameter of 6.050 Å (SI, section 4). For each titration, 1 ml of NCs solution was injected to the ITC sample cell and the ligand solution was loaded to the 250 µl ITC syringe. The surface sites and the ligands concentration were adjusted in order to produce high quality titration curves. At each injection step 5 µl of ligands solution was injected to the cell and the heat flow was measured for 600-800 seconds during which the system returned to equilibrium. All ITC thermograms and exchange-model fitted titration curves, including detailed derivation of the single-site model, are presented in the SI, section 4.

**Calculations of thermodynamic parameters for mixing and ligand shell structure:** The ITC-extracted thermodynamic parameters were fitted to the sum of the linear combination and the calculated $\Delta G_{mix}$, $\Delta H_{mix}$ and $\Delta S_{mix}$ using the thermodynamic integration method. The binary shell structure was simulated based on a lattice model according to the interaction free energy parameter between the ligands, $\chi_G$. Generally, paring two different ligands is allowed with the Boltzmann probability $\exp(\Delta\mathcal{G}/RT)$, where $\Delta\mathcal{G}$ is derived directly from $\chi_G$ and the change in the ligand's nearest neighbors due to the pairing process. In the first step of the fitting procedure, $\Delta G_{mix}$ is evaluated, while the temperature dependence of $\chi_G$ allowed calculating also $\Delta H_{mix}$ and $\Delta S_{mix}$. Further model details are in SI, section 6.

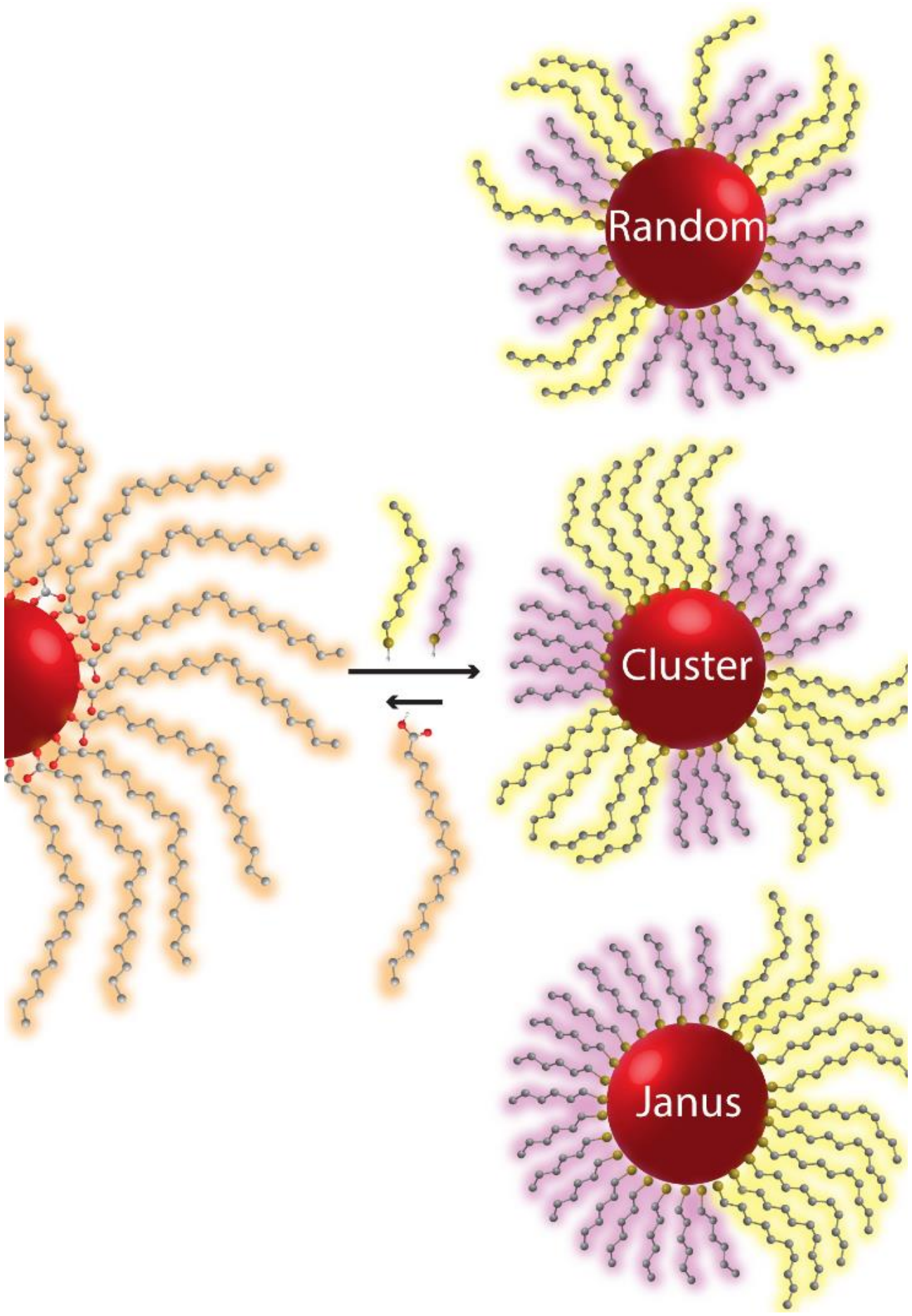

**Scheme 1:** The investigated ligand exchange reaction between oleate coated CdSe NCs and a binary mixture of thiolated ligands. Shown are the possible organizations of the thiolated ligand shells, classified into random packing (involving full mixing of the two ligands), clustering (where segregated patches of the different ligands form), and Janus behavior (where the two ligand types occupy separate regions on the NC surface).

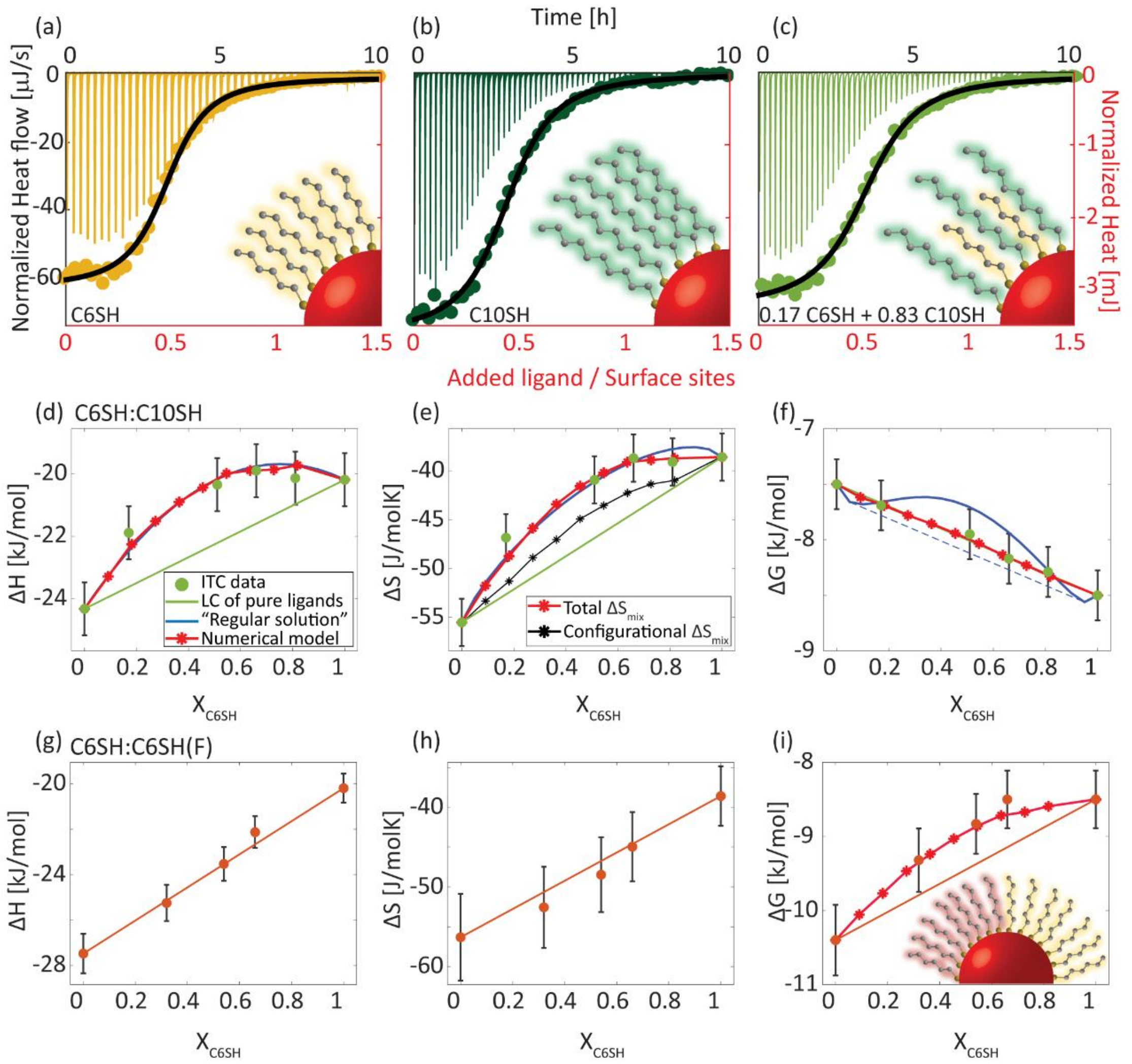

**Figure 1:** ITC results for ligand exchange reaction of oleate coated CdSe NCs (d=3.0nm) with pure and mixed ligands (a-c) Real-time ITC thermograms (black axis) and the corresponding titration curves and fittings (dots, black line respectively, red axis) for the exchange with (a) pure hexanethiol (C6SH), (b) pure decanethiol (C10SH), and (c) a $C6SH_{0.17}C10SH_{0.83}$ binary composition. Inset: Illustration of the yielded systems. (d-f) The extracted thermodynamic parameters: (d) enthalpy, (e) entropy, and (f) Gibbs free energy, for the exchange with pure C6SH ($X_{C6SH}$=1), C10SH ($X_{C6SH}$=0) and their binary composition (0<$X_{C6SH}$<1) (green dots). Solid green line represents the calculated linear combination (LC) of the pure ligands. Solid blue line represents fitting to the regular solution model. Red asterisk and line represent the numerical model.

Black asterisk and line represent the calculated configurational entropy. (g-i) The extracted thermodynamic parameters: (g) enthalpy, (h) entropy, and (i) Gibbs free energy, for the exchange reaction with pure C6SH, fluorinated hexanethiol (C6SH(F)) and their binary compositions (orange dots). Solid orange line represents the calculated linear combination of the pure ligands. Red asterisk and line represent the numerical model. Inset: Illustration of the surface ligand organization.

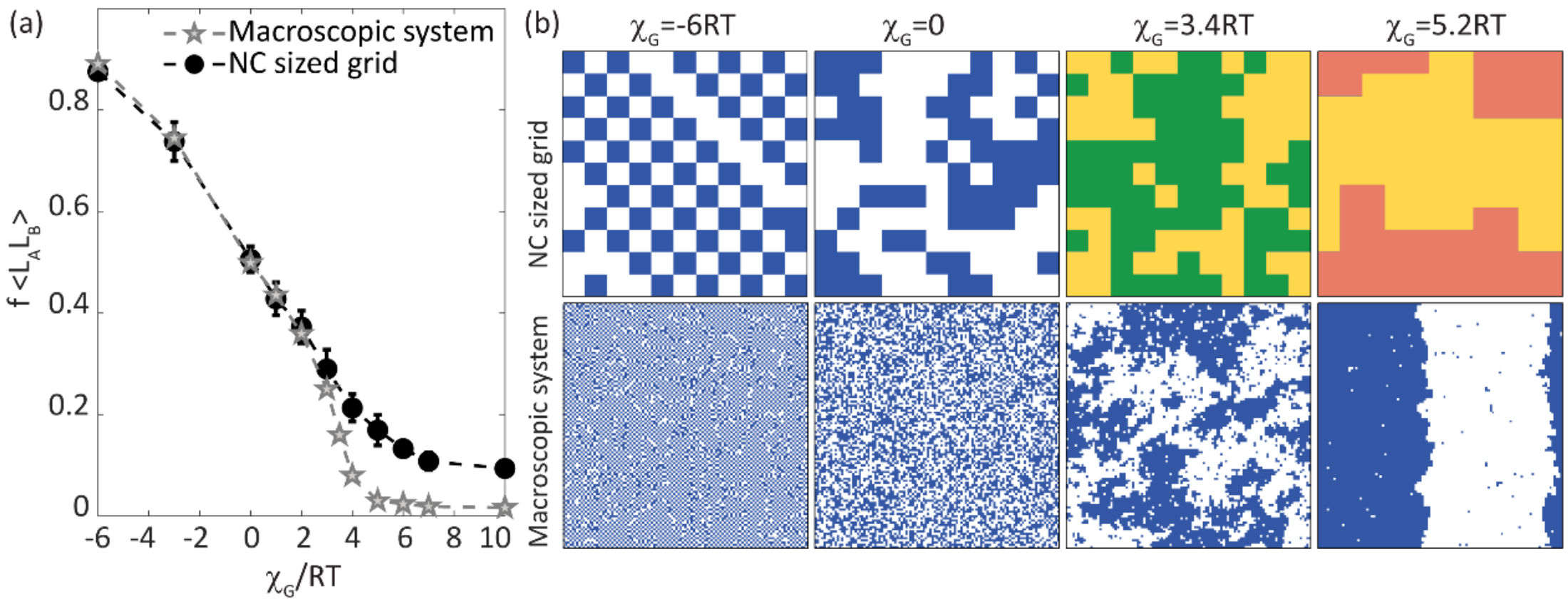

**Figure 2:** Comparison of patterning of ligands in macroscopic and NC-sized simulations. (a) Average fraction of different ligands pairs ($L_AL_B$), as a function of the total interaction parameter, $\chi_G$, (normalized to the thermal energy RT). The molar fraction of surface ligands is X=0.5. (b) Ligand shell organization for several $\chi_G$ values in the NC sized grid (top) and the macroscopic system (bottom). Ligand shell patterning corresponding to interaction values for C6SH:C10SH and C6SH:C6SH(F) are presented in yellow & green and yellow & orange, respectively.

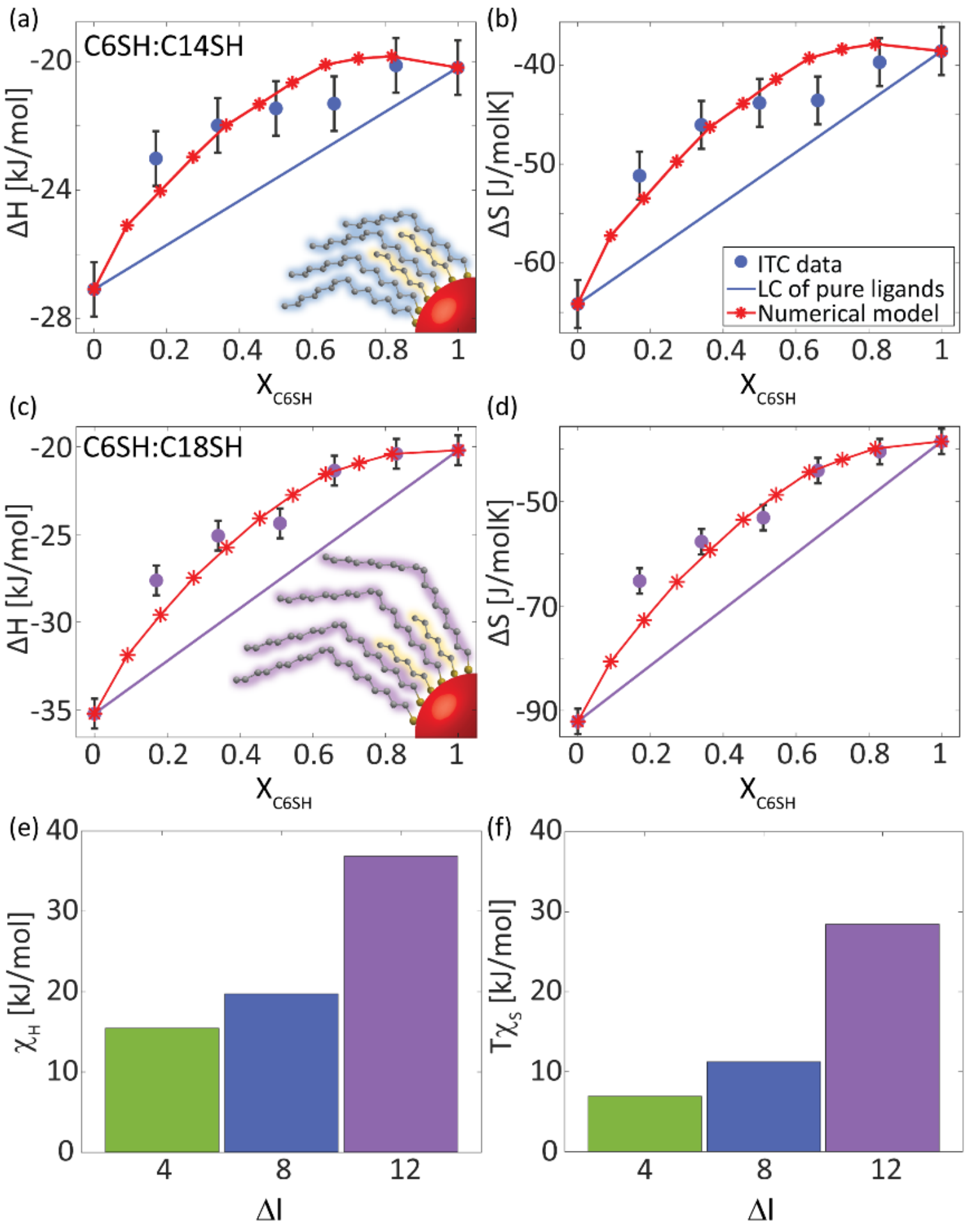

**Figure 3:** ITC-extracted (a, c) Enthalpy and (b,d) entropy for the exchange reaction of oleate coated CdSe NCs (d=3.0nm) with (a-b) C6SH:C14SH and (c-d) C6SH:C18SH binary compositions ($0<X_{C6SH}<1$, blue and purple dots, respectively). Blue and purple lines represent the linear combination (LC) of the pure ligands as a reference, respectively. Red asterisks and lines represent a fit to the numerical model. Inset: Illustration of the attached ligands. (e-f) Summary of the fitting-extracted (e) enthalpic and (f) entropic interaction parameters and their ligand length (Δl) dependency.

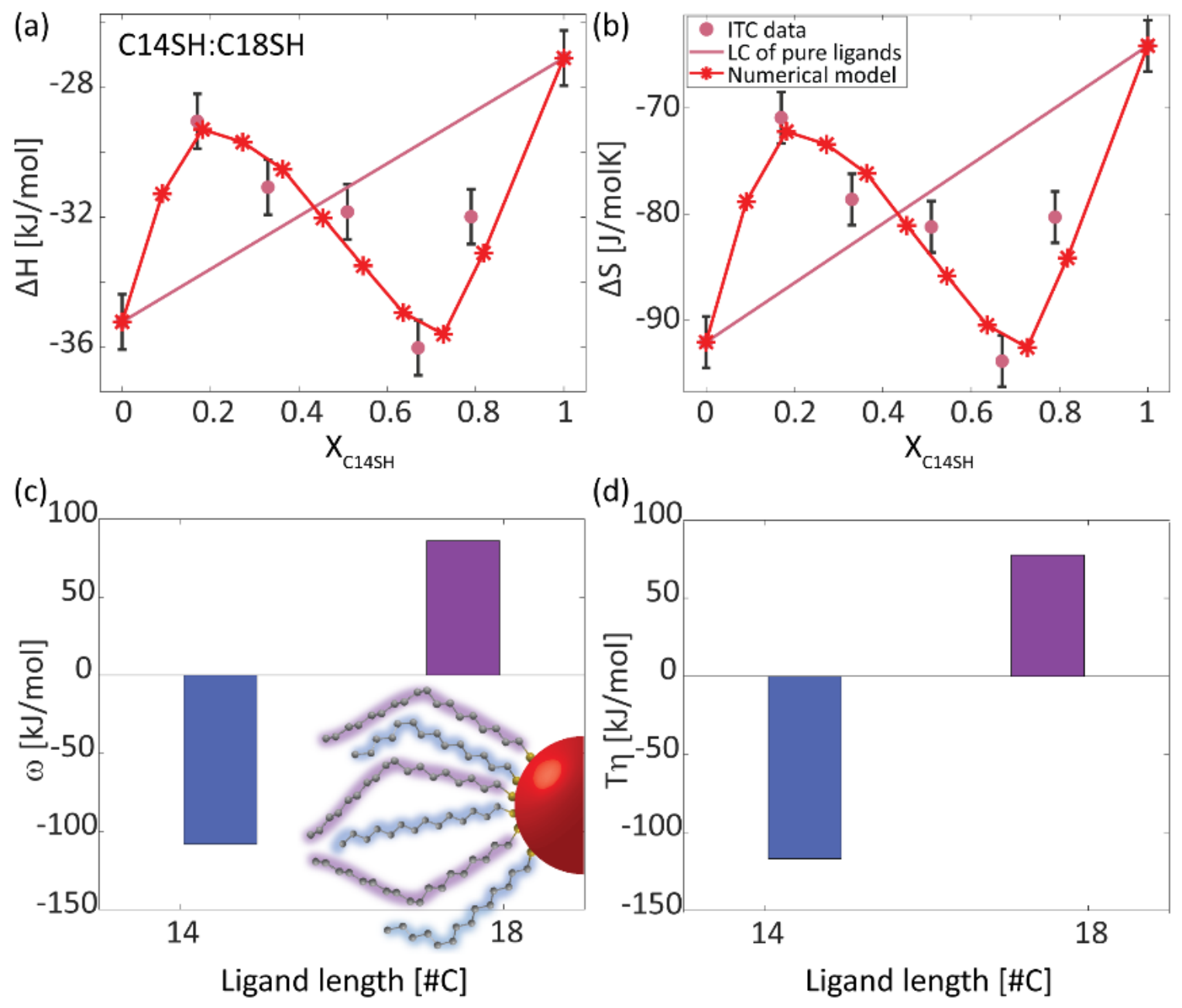

**Figure 4:** (a) ITC-extracted Enthalpy and (b) entropy for the exchange reaction of oleate coated CdSe NCs (d=3.0nm) with a mixture of C14SH and C18SH at different ratios (pink dots). The parameters of the pure ligands and their linear combination (LC) are presented for reference (pink line). The fitting of the ITC data with a numerical model is presented in red asterisks and lines (c) Fitting-extracted enthalpic and (d) entropic interaction parameters coefficient for both ligands. Inset: Illustration of the attached ligands.

ASSOCIATED CONTENT

Supporting Information (PDF) containing the ITC raw data thermograms, titration curves and fitting to the exchange model, surface analysis via TGA, as well as a detailed explanation of the mixing models discussed in the main text, is available free of charge.

AUTHOR INFORMATION

**Corresponding Authors**

Uri Banin*

The Institute of Chemistry and The Center for Nanoscience and Nanotechnology, The Hebrew University of Jerusalem, Jerusalem 9190401, Israel.

E-mail: Uri.Banin@mail.huji.ac.il

Daniel Harries*

The Institute of Chemistry, The Fritz Haber Center, and The Center for Nanoscience and Nanotechnology, The Hebrew University of Jerusalem, Jerusalem 9190401, Israel.

E-mail: Daniel.Harries@mail.huji.ac.il

**Author contributions**

The manuscript was written through contributions of all authors. All authors have given approval to the final version of the manuscript.

**Funding sources**

European Research Council (ERC) under the European Union's Horizon 2020 research and innovation program (grant agreement No [741767], CoupledNC).

**Notes**

The authors declare no competing financial interest.

ACKNOWLEDGMENT

The research leading to these results has received financial support from the European Research Council (ERC) under the European Union's Horizon 2020 research and innovation program (grant agreement No [741767], CoupledNC). O.E acknowledges support from the Hebrew university center for nanoscience and nanotechnology. U.B. thanks the Alfred & Erica Larisch memorial chair.

`

# Supporting Information

# Spontaneous Patterning of Binary Ligand Mixtures on CdSe Nanocrystals: from Random to Janus Packing

*Orian Elimelech,*[a] *Meirav Oded,*[a] *Daniel Harries,*[a,b]* *and Uri Banin*[a]*

a The Institute of Chemistry and The Center for Nanoscience and Nanotechnology, The Hebrew University of Jerusalem, Jerusalem 9190401, Israel.

b The Fritz Haber Center, The Hebrew University of Jerusalem, Jerusalem 9190401, Israel.

*E-mail: Uri.Banin@mail.huji.ac.il

Daniel.Harries@mail.huji.ac.il

`

# Table of contents:

## 1. CdSe NCs synthesis

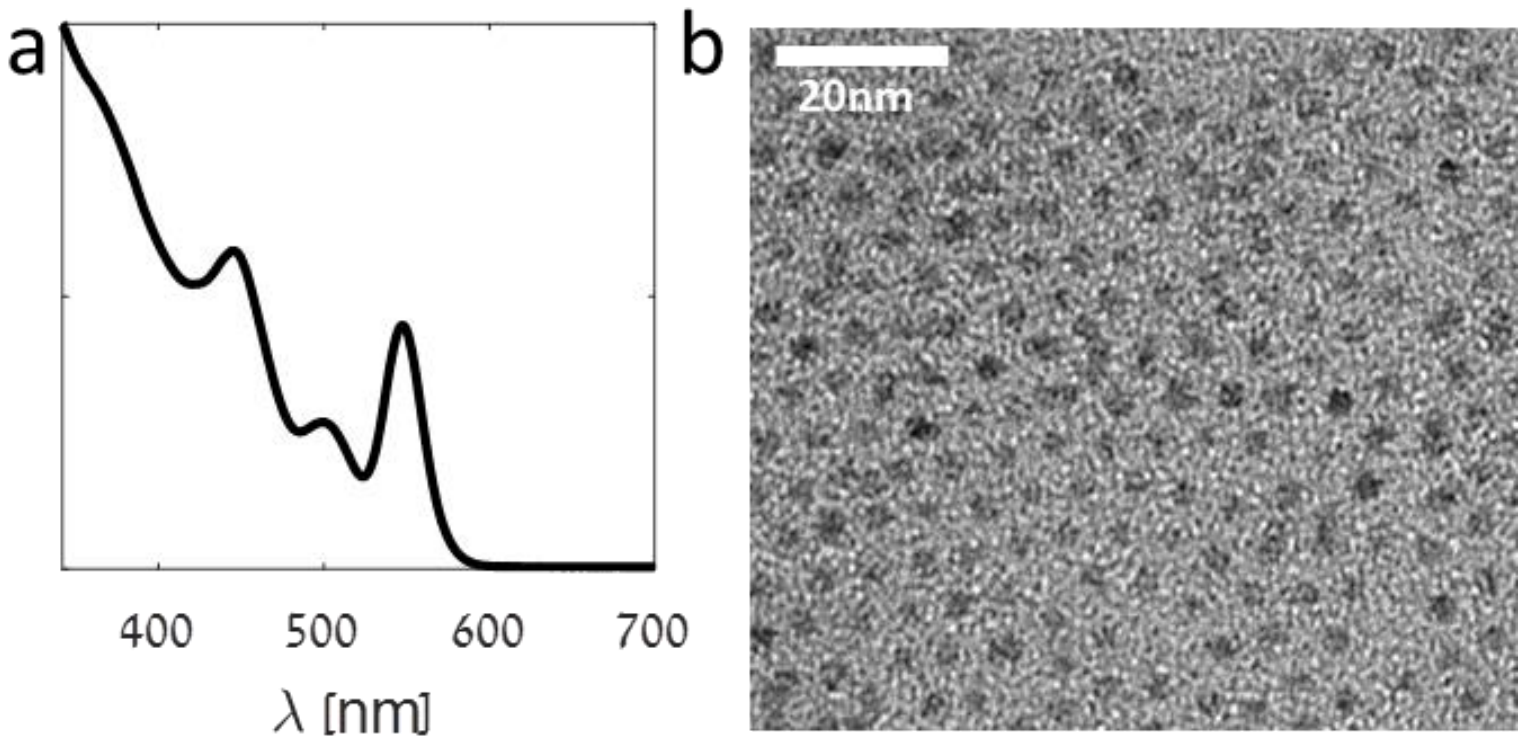

***Figure S1.*** *(a) Absorption spectra and (b) TEM image of as-synthesized oleate coated CdSe NCs.*

## 2. n-Alkylthiols reduction

To avoid inaccuracies in the ligands concentration, derived by S-S chain coupling bonds, the purchased ligands were reduced prior to use, as was discussed in detailed elsewhere.[1]. Briefly, two equivalents of $NaBH_4$ powder were added to a solution of alkylthiol in ethanol and TDW (1:4). After 12 hours of stirring at room temperature, the solution was extracted with chloroform (3x75 ml portions). Following that, the unified organic phase was dried over $MgSO_4$ , filtered and then evaporated under vacuum in order to separate between the chloroform and the reduced alkylthiol. The reduced ligands were kept under an inert atmosphere with no exposure to UV light for future use. The yield of the reduction procedure is 70%, to give a final product with no more than 5% of disulfide.

## 3. Surface sites calculation

As described in our previous studies,[1,2] the number of Cd surface sites ($N_{surface}$) was calculated based on a simple spherical model for the NCs with a lattice parameter of a=6.050Å (zinc blende). We assumed a uniform zinc blende CdSe layer on the surface, hence the number of Cd surface sites is

(S1) $$N_{surface} = N_{total} - N_{internal-sphere}$$

where $N_{total}$ is the total number of Cd atoms in the NC and $N_{internal\text{-}sphere}$ is the number of core Cd. $N_{total}$ was calculated considering the volume of a spherical NC with a radius $R_{NC}$, the density of CdSe ($\rho_{CdSe}$) and its molar mass ($Mw_{cdSe}$):

(S2) $$N_{total} = \frac{\frac{4}{3}\pi R_{NC}^3 \cdot \rho_{CdSe}}{Mw_{CdSe}} \cdot N_A$$

`

$N_{internal\text{-}sphere}$ was calculated in a similar way to $N_{total}$ excluding the outer layer of the surface Cd:

(S3) $$N_{internal-sphere} = \frac{\frac{4}{3}\pi\left(R_{NC}-\frac{a}{2}\right)^3 \cdot \rho_{CdSe}}{Mw_{CdSe}} \cdot N_A$$

For the investigated d=3.0 nm NC, 127 Cd surface sites are expected.

The results were compared with a pyramidal model for zinc blend CdSe NCs with four exposed (111) facets. The height of the pyramid, *h,* was taken as the calculated diameter of the NC, hence the edge length, *c*, is:

(S4) $$c = \sqrt{\frac{3}{2}}h$$

The Cd atoms are spaced on the edge according to the nearest-neighbor distance, *d*, of the unit cell:

(S5) $$d = \frac{\sqrt{2}}{2}a$$

Hence, the length of an edge, *c*, containing *N* atoms is:

(S6) $$c = \frac{1}{\sqrt{2}}a(N-1)$$

Using eq. (S4) in eq. (S6), we can determine the number of atoms on the edge:

(S7) $$N = \frac{h\sqrt{3}}{a} + 1$$

Since the zinc blend NCs are actually a truncated pyramid, the atoms of the outer edges were removed, and the new faces lost one atoms per line, per side. Therefore, the number of Cd surface atoms on a single face, $N_{face}$, with a base containing *(N-2)* atoms is calculated by:

(S8) $$N_{face} = \sum_{q=1}^{N-2} q = \frac{(n-2)(n-2+1)}{2}$$

By using eq. (S7) in eq. (S8) and multiplying it by 4 (for the four faces), we find the total number of Cd surface atoms in all four faces as a function of the pyramid height:

(S9) $$N_{surface} = 2\left(\frac{h\sqrt{3}}{a} + 1\right)^2 - 6\left(\frac{h\sqrt{3}}{a} + 1\right) + 4$$

For the investigated d=3.0nm NC, 130 Cd surface sites are expected.

In addition, an atomistic model was also considered to verify the suggested models for surface sites. A semi-spherical NC was simulated from the bulk zinc blend CdSe crystal structure by removing atoms located beyond a distance that is greater than the desired radius (Figure S2). The remaining atoms resulted in a non-stoichiometric ratio between Cd and Se atoms. The atoms in the outer layer were considered as surface sites. To account for the range of error in the size estimation, results for NC diameters of 2.8, 3.0 and 3.2 nm are presented in Table S1. For an average d=3.0±0.2 nm NC (similar to the experimentally extracted distribution of the investigated NCs,), 128±47 Cd surface sites are expected with an average Cd:Se ratio of

1.1±0.1. The non-stoichiometric ratio we find is consistent with previous reports on CdSe NCs.[3]

All presented models give similar average numbers of Cd surface sites.

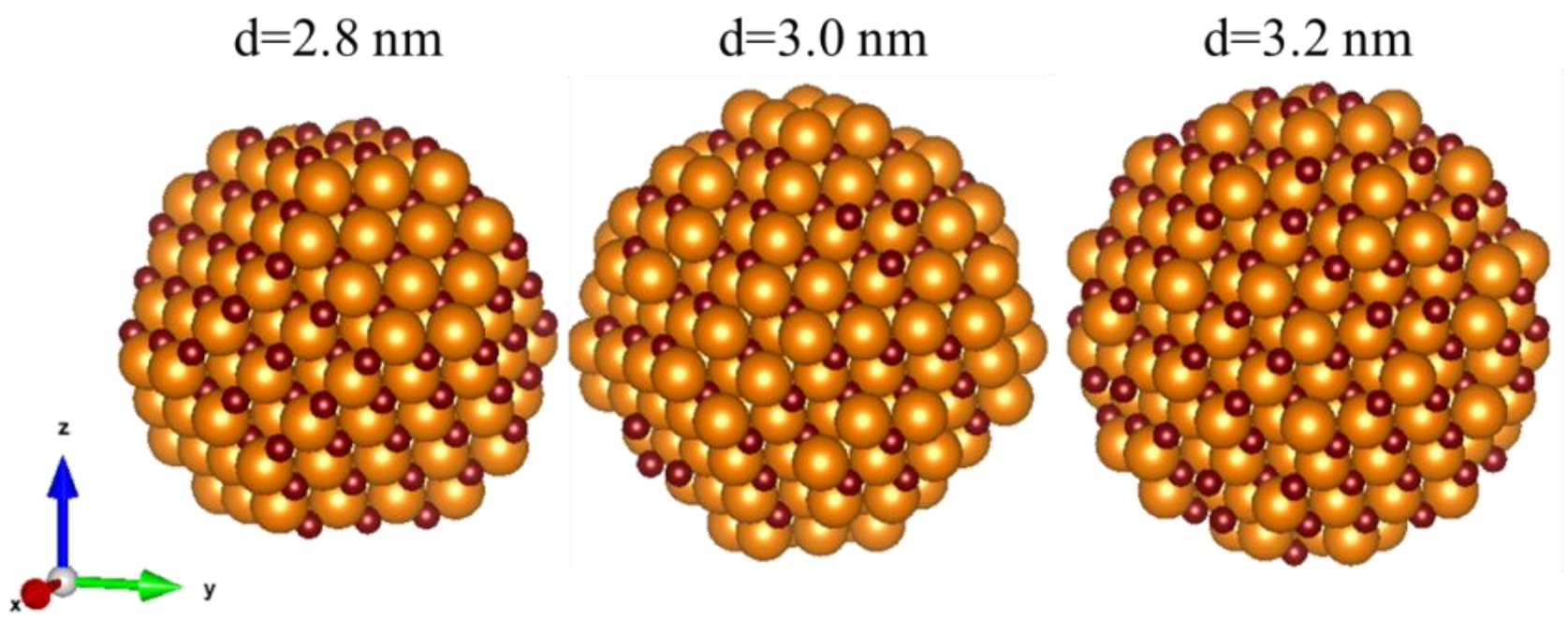

***Figure S2.*** *Simulated atomistic model for zinc blend CdSe NC of diameters 2.8, 3.0 and 3.2 nm. Cd and Se atoms are colored brown and orange respectively.*

| **NC diameter** | **Cd atoms** | **Cd:Se ratio** | **Surface Cd** | **Surface Se** |
|---|---|---|---|---|
| **2.8** | 201 | 1.14 | 114 | 96 |
| **3.0** | 225 | 0.94 | 90 | 100 |
| **3.2** | 321 | 1.16 | 180 | 124 |

***Table S1.*** *Summary of the simulated atomistic model data for zinc blend CdSe NC of diameters 2.8, 3.0 and 3.2 nm, as presented in Figure S2.*

`

# 4. ITC measurements and analysis

4.1. Derivation of a single-site ligand exchange model

As described in our previous studies,[1,2] the ligand exchange model is based on the well-known "single set of independent binding sites" model.[4] We modified the known “binding” model, which considers only the attachment of the new ligand, in order to take into account also the detachment of the native ligand. This “exchange” model is necessary for ligand exchange reactions since both processes, detachment and attachment of the ligands, release heat which is measured by the ITC instrument.

The ligand exchange reaction between the native ligand $L'$ and the exchanged ligand $L$ for a single surface site $M$ is

(S10) $ML' + L \rightleftharpoons ML + L'$

The equilibrium constant is defined as

(S11) $K = \frac{[ML][L']}{[ML'][L]}$

Assuming that each native ligand $L'$ is exchanged with a single new ligand $L$ and no free $L'$ is present initially[5] (supported by TGA data, see in the next section), we get:

(S12) $[ML] = [L']$

While for the exchanged ligands:

(S13) $[L] = [L]_0 - [ML]$

And for all surface sites:

(S14) $[ML'] = n[M]_0 - [ML]$

In the previous equations, $[M]_0$ is the total number of surface sites on the NC (based on a spherical model, as explained before), *n* is the ratio between the actually exchanged ligands and the available surface sites (i.e., the reaction stoichiometry coefficient), and $[L]_0$ is the total added exchanged ligand.

Given the expressions above, the equilibrium constant can be written as:

(S15) $K = \frac{[ML]^2}{(n[M]_0 - [ML])([L]_0 - [ML])}$

We define $\theta$ as the NC surface coverage, and hence,

(S16) $[ML] = n[M]_0\theta.$

Given eq. (S12), eq. (S15) can be rewritten as

(S17) $0 = \theta^2 - \theta\left(\frac{K}{K-1}\right)\left(1 + \frac{[L]_0}{n[M]_0}\right) + \left(\frac{K}{K-1}\right)\left(\frac{[L]_0}{n[M]_0}\right)$

During an ITC experiment, we measure the total amount of heat released per injection of ligand, which is correlated with the enthalpy change of the reaction

`

(S18) $Q_{total} = \theta n[M]_0 V_{cell} \Delta H$

By using the solution for the quadratic equation(S17), eq. (S18) can be written as

(S19) $$Q_{total} = \frac{n[M]_0 V_{cell}\Delta H}{2}\left[\left(\frac{K}{K-1}\right)\left(1+\frac{[L]_0}{n[M]_0}\right) - \sqrt{\left(\frac{K}{K-1}\right)^2\left(1+\frac{[L]_0}{n[M]_0}\right)^2 - 4\left(\frac{K}{K-1}\right)\left(\frac{[L]_0}{n[M]_0}\right)}\right]$$

and the heat released per injection of ligand is

(S20) $$\frac{dQ_{tot}}{d[L]_0} = \frac{V_{cell}\Delta H}{2}\left(\frac{K}{K-1}\right)\left[1 - \frac{\frac{[L]_0}{n[M]_0}+\frac{2-K}{K}}{\sqrt{1+\left(\frac{[L]_0}{n[M]_0}\right)^2+\left(\frac{[L]_0}{n[M]_0}\right)\left(\frac{4-2K}{K}\right)}}\right]$$

where

(S21) $$d[L]_0 = \frac{V_{injection}[L]_{syringe}}{V_{cell}}$$

By implanting eq. (S21) into eq. (S20), we get the final equation for fitting

(S22) $$dQ_{tot} = \frac{V_{injection}[L]_{syringe}\Delta H}{2}\left(\frac{K}{K-1}\right)\left[1 - \frac{\frac{[L]_0}{n[M]_0}+\frac{2-K}{K}}{\sqrt{1+\left(\frac{[L]_0}{n[M]_0}\right)^2+\left(\frac{[L]_0}{n[M]_0}\right)\left(\frac{4-2K}{K}\right)}}\right]$$

The other thermodynamics parameters $\Delta G$ and $\Delta S$ are calculated by using the known thermodynamics relations

(S23) $\Delta G = -RT \ln K$

(S24) $\Delta S = \frac{\Delta H - \Delta G}{T}$

`

## 4.2. Experimental data and fitting

All fittings were done in NanoAnalyze Software v 3.10.0 (TA instrument).

### *4.3.1. Error analysis:*

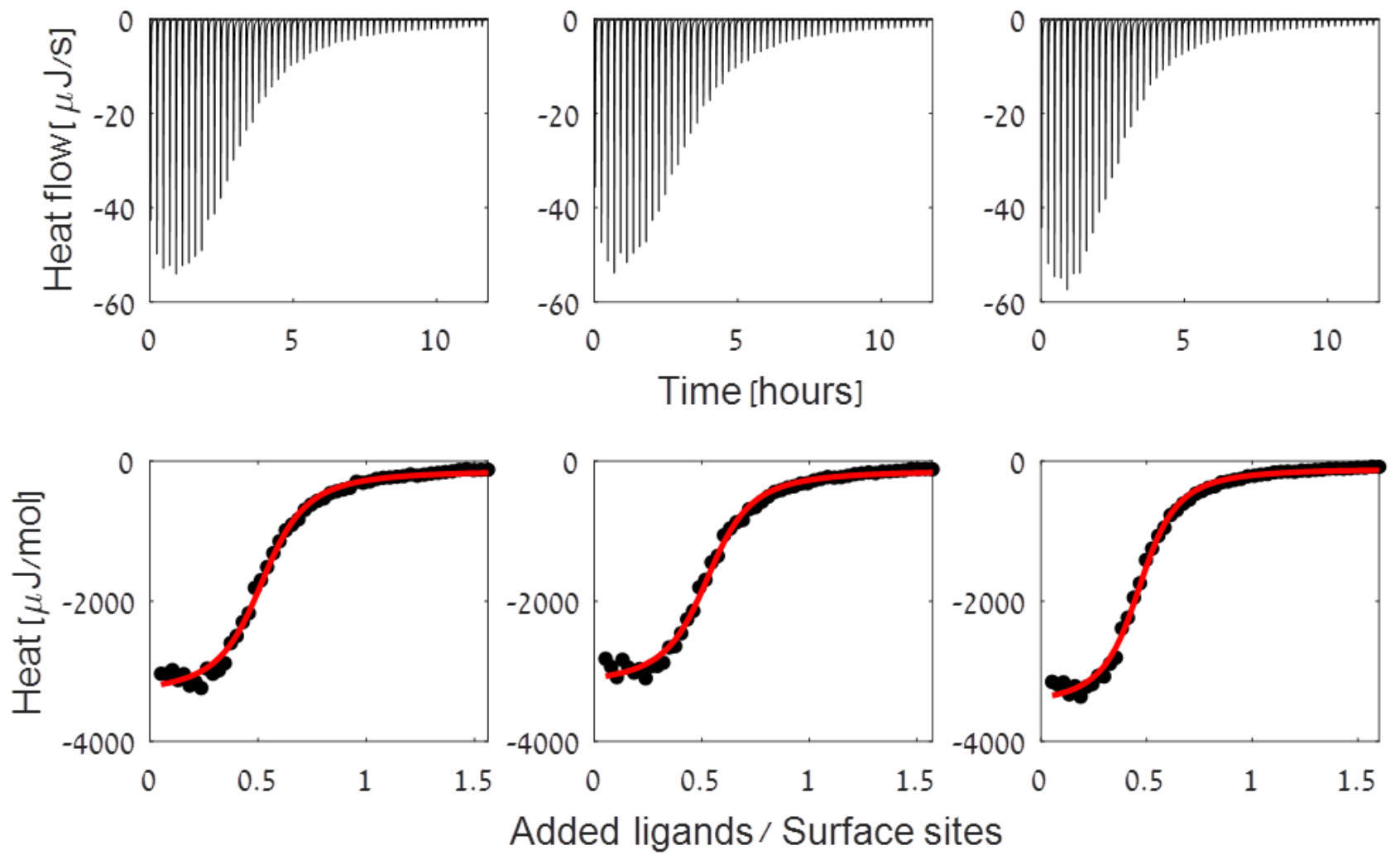

***Figure S3.*** *Real-time thermograms and the corresponding titration curves with their model fitting for the ligand exchange reaction of oleate-coated CdSe NCs with an equimolar mixture (1:1) of C6SH and C14SH at 303K.*

| | Ligand [mM] | Surface sites [mM] | ΔH [kJ/mol] | ΔS [J/molK] | ΔG [kJ/mol] | n |
|---|---|---|---|---|---|---|
| **Exp. 1** | 30 | 5.7±0.2 | -23.2±0.7 | -48±5 | -8.5±0.4 | 0.49±0.03 |
| **Exp. 2** | 29 | 5.6±0.2 | -22.4±0.7 | -47±5 | -8.0±0.4 | 0.55±0.03 |
| **Exp. 3** | 30 | 5.8±0.2 | -21.5±0.7 | -44±4 | -8.2±0.4 | 0.56±0.03 |

***Table S2.*** *The thermodynamic parameters extracted from the model fit of the titration curves for the ligand exchange reaction of oleate-coated CdSe NCs with an equimolar mixture (1:1) of C6SH and C14SH at 303K, presented in Figure S3. Errors were calculated based on the quality of the fitting.*

Errors of the extracted thermodynamics parameters were determined by the quality of the fitting. In addition, we considered the reproducibility of the measurement by performing the same experiments three times and calculating the standard deviation of each parameter.

The error in the surface site's concentration was calculated by a triple measurement of the absorption.

### 4.3.2. ITC data and analysis for ligand exchange with pure alkylthiols

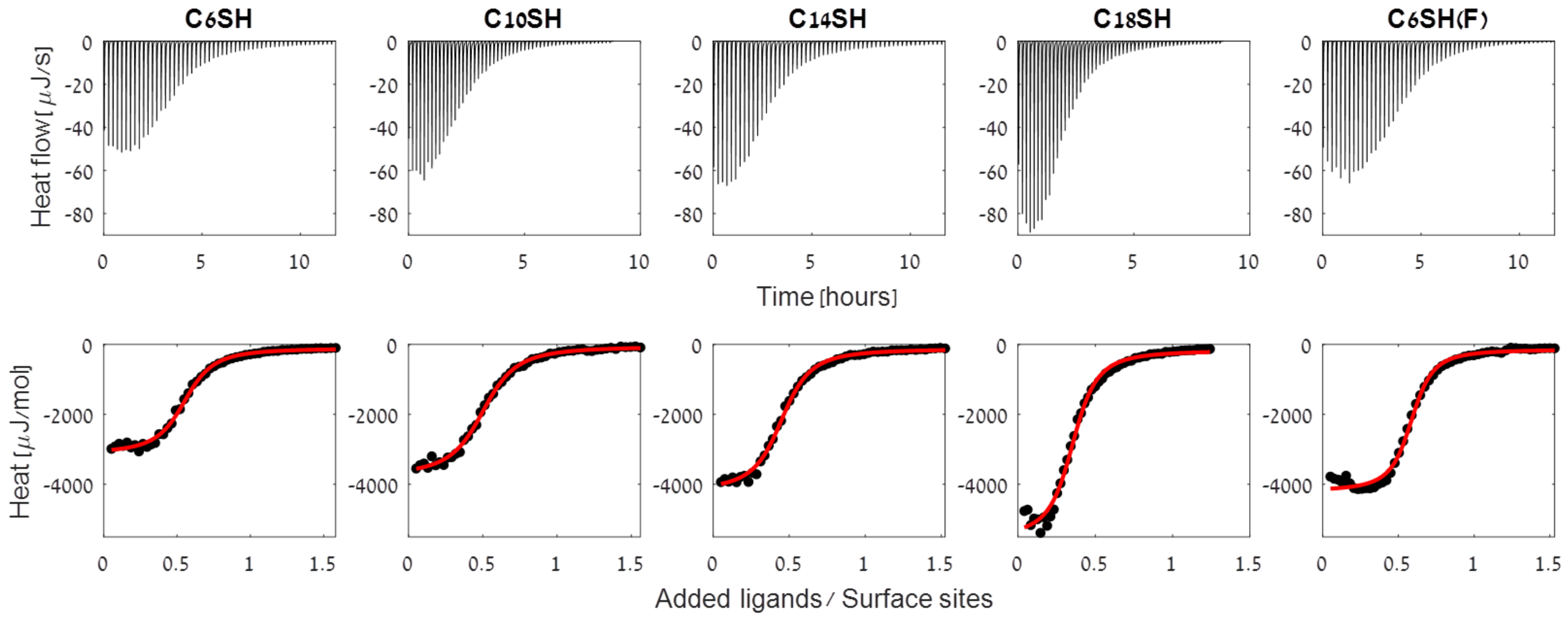

***Figure S4.*** *Real-time thermograms and the corresponding titration curves with their model fittings (red) for the ligand exchange reaction of oleate-coated CdSe NCs with pure alkylthiols at 303K.*

| Ligand | Ligand [mM] | Surface sites [mM] | ΔH [kJ/mol] | ΔS [J/molK] | ΔG [kJ/mol] | n |
|---|---|---|---|---|---|---|
| **1-Hexanethiol (C6SH)** | 31 | 5.9 | -20.2 | -39 | -8.5 | 0.58 |
| **1-Decanethiol (C10SH)** | 30 | 5.8 | -24.3 | -56 | -7.5 | 0.54 |
| **1-Tetradecanethiol (C14SH)** | 31 | 6.1 | -27.1 | -64 | -7.6 | 0.48 |
| **1- Octadecanethiol (C18SH)** | 23 | 5.7 | -35.2 | -92 | -7.3 | 0.46 |
| **1H,1H,2H,2H-Perfluoro-1-hexanethiol (C6SH(F))** | 30 | 5.9 | -27.5 | -56 | -10.4 | 0.59 |

***Table S3.*** *A table summarizing the thermodynamic parameters extracted from the single-site model fit of the titration curves for the ligand exchange reaction of oleate-coated CdSe NCs with pure alkylthiol at 303K, presented in Figure S4.*

### 4.3.3. ITC data and analysis for ligand exchange with a mixture of alkylthiols

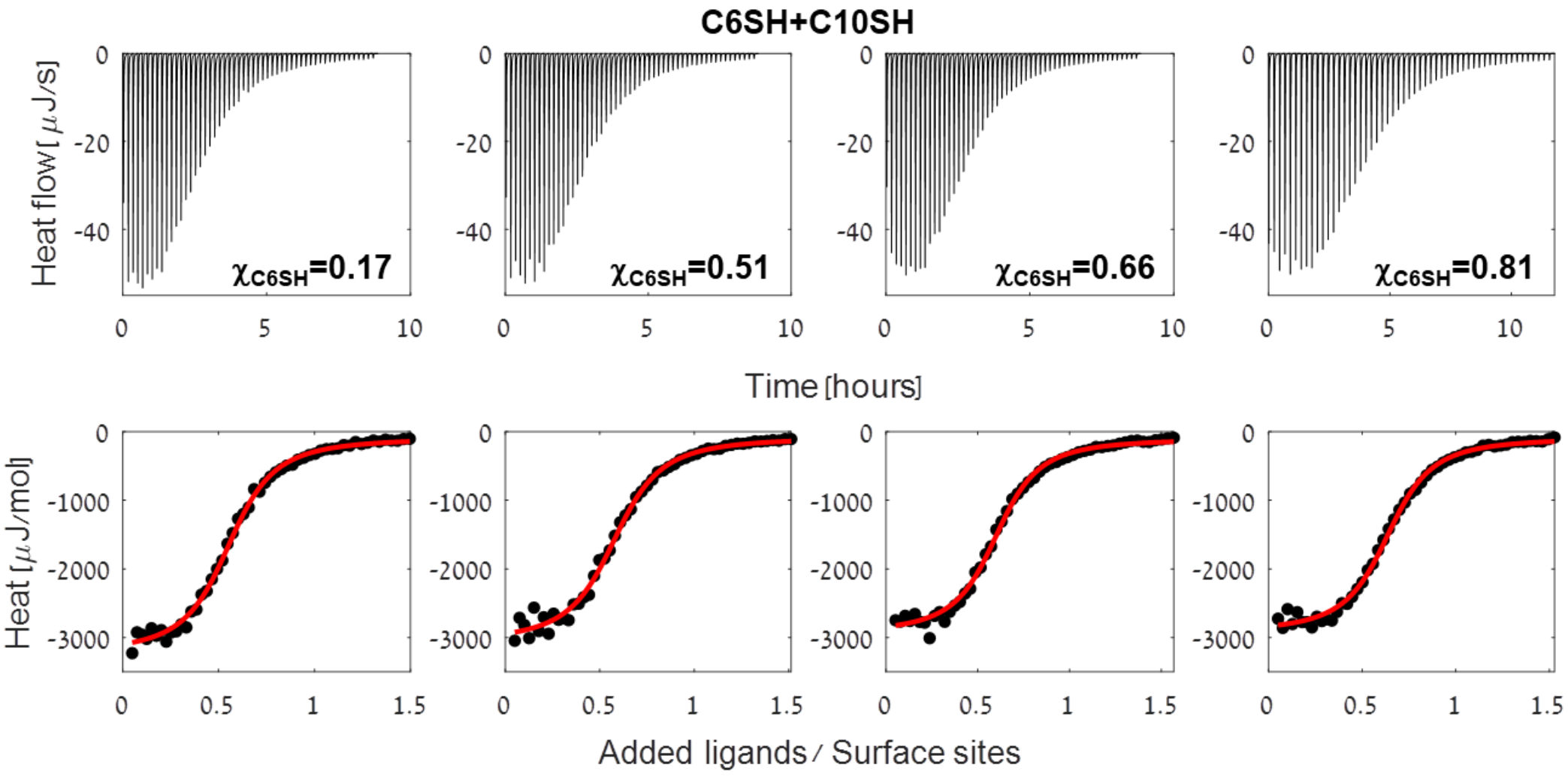

***Figure S5.*** *Real-time thermograms and the corresponding titration curves with their model fittings (red) for the ligand exchange reaction of oleate-coated CdSe NCs with different mixtures of C6SH and C10SH at 303K.*

| χ(C6SH) | Ligand [mM] | Surface sites [mM] | ΔH [kJ/mol] | ΔS [J/molK] | ΔG [kJ/mol] | n |
|---|---|---|---|---|---|---|
| **0.17** | 29 | 5.9 | -21.9 | -47 | -7.7 | 0.59 |
| **0.51** | 30 | 6.1 | -20.4 | -41 | -7.9 | 0.62 |
| **0.66** | 28 | 5.5 | -19.9 | -38 | -8.2 | 0.63 |
| **0.81** | 29 | 5.7 | -20.1 | -39 | -8.3 | 0.67 |

***Table S4.*** *A table summarizing the thermodynamic parameters extracted from the single-site model fit of the titration curves for the ligand exchange reaction of oleate-coated CdSe NCs with different mixtures of C6SH and C10SH at 303K, presented in Figure S5.*

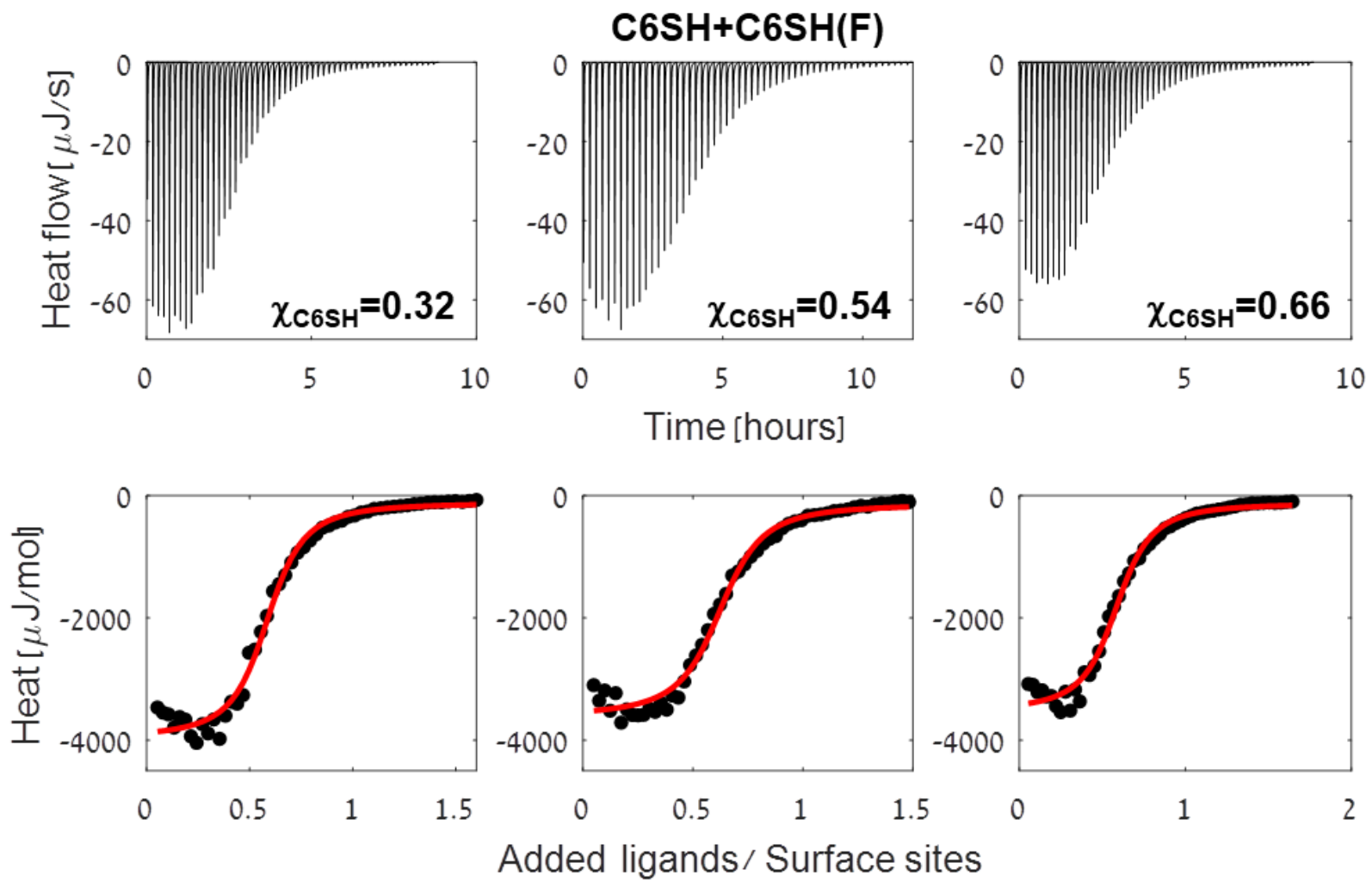

***Figure S6.*** *Real-time thermograms and the corresponding titration curves with their model fittings (red) for the ligand exchange reaction of oleate-coated CdSe NCs with different mixtures of C6SH and C6SH(F) at 303K.*

| χ(C6SH) | Ligand [mM] | Surface sites [mM] | ΔH [kJ/mol] | ΔS [J/molK] | ΔG [kJ/mol] | n |
|---|---|---|---|---|---|---|
| **0.32** | 31 | 5.9 | -25.3 | -52 | -9.3 | 0.60 |
| **0.54** | 30 | 6.1 | -23.5 | -48 | -8.8 | 0.64 |
| **0.66** | 30 | 5.4 | -22.1 | -45 | -8.5 | 0.61 |

***Table S5.*** *A table summarizing the thermodynamic parameters extracted from the single-site model fit of the titration curves for the ligand exchange reaction of oleate coated CdSe NCs with different mixtures of C6SH and C6SH(F) at 303K, presented in Figure S6.*

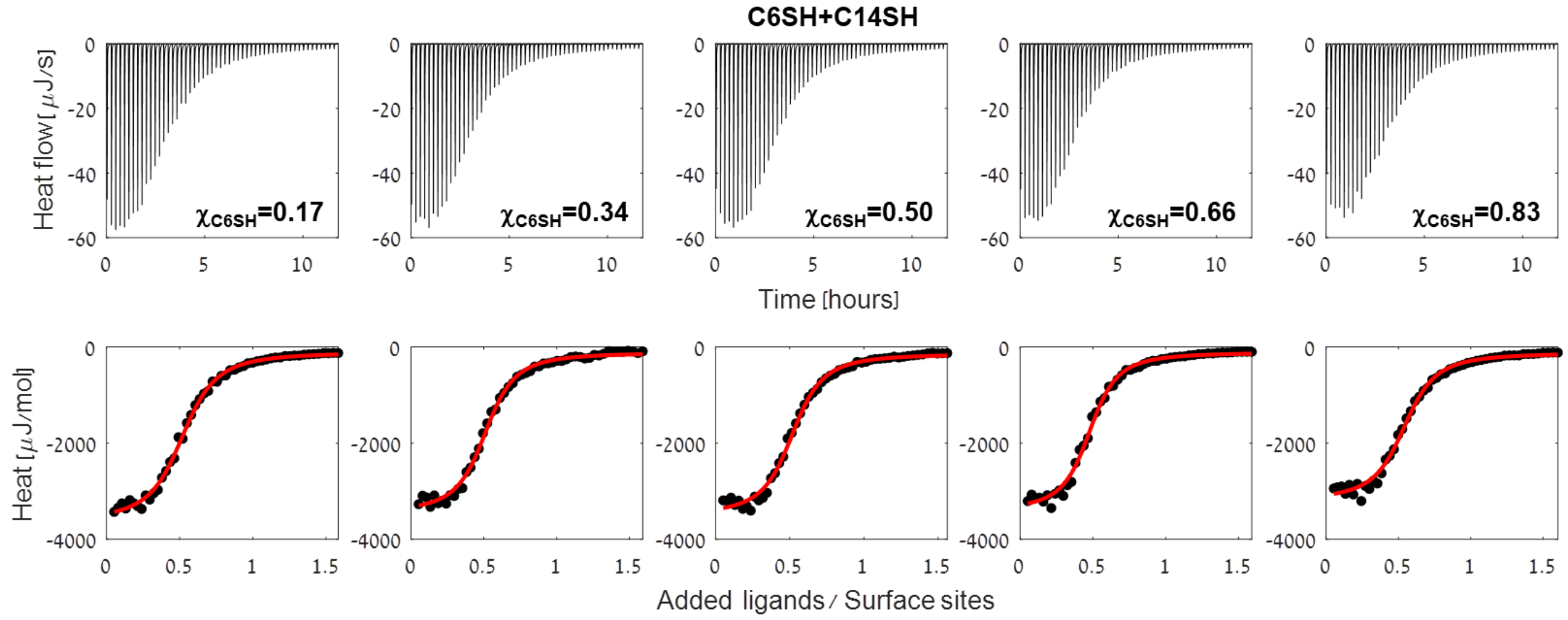

***Figure S7.*** *Real-time thermograms and the corresponding titration curves with their model fittings (red) for the ligand exchange reaction of oleate coated CdSe NCs with different mixtures of C6SH and C14SH at 303K.*

| **χ(C6SH)** | **Ligand [mM]** | **Surface sites [mM]** | **ΔH [kJ/mol]** | **ΔS [J/molK]** | **ΔG [kJ/mol]** | **n** |
|---|---|---|---|---|---|---|
| **0.17** | 31 | 5.8 | -23.0 | -51 | -7.5 | 0.56 |
| **0.34** | 30 | 5.8 | -22.0 | -46 | -8.0 | 0.54 |
| **0.50** | 30 | 5.8 | -21.5 | -44 | -8.2 | 0.56 |
| **0.66** | 30 | 5.8 | -21.3 | -43 | -8.1 | 0.51 |
| **0.83** | 30 | 5.7 | -20.1 | -39 | -8.1 | 0.57 |

***Table S6.*** *A table summarizing the thermodynamic parameters extracted from the single-site model fit of the titration curves for the ligand exchange reaction of oleate-coated CdSe NCs with different mixtures of C6SH and C14SH at 303K, presented in Figure S7.*

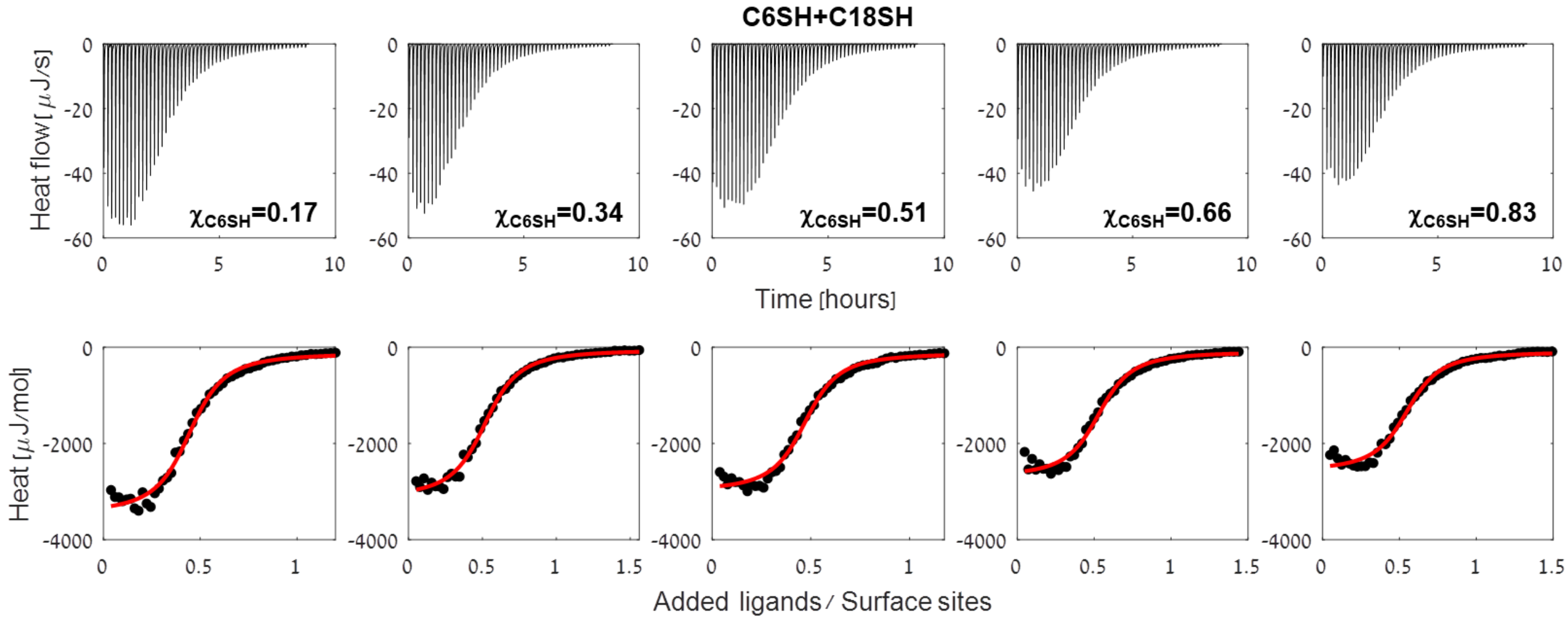

***Figure S8.*** *Real-time thermograms and the corresponding titration curves with their model fittings (red) for the ligand exchange reaction of oleate-coated CdSe NCs with different mixtures of C6SH and C18SH at 303K.*

| **χ(C6SH)** | **Ligand [mM]** | **Surface sites [mM]** | **ΔH [kJ/mol]** | **ΔS [J/molK]** | **ΔG [kJ/mol]** | **n** |
|---|---|---|---|---|---|---|
| **0.17** | 25 | 6.3 | -27.6 | -635 | -7.8 | 0.47 |
| **0.34** | 31 | 6.0 | -25.1 | -58 | -7.6 | 0.56 |
| **0.51** | 25 | 6.3 | -24.4 | -53 | -8.2 | 0.49 |
| **0.66** | 29 | 6.0 | -21.3 | -44 | -8.0 | 0.56 |
| **0.83** | 30 | 6.1 | -20.4 | -40 | -8.1 | 0.58 |

***Table S7.*** *A table summarizing the thermodynamic parameters extracted from the single-site model fit of the titration curves for the ligand exchange reaction of oleate-coated CdSe NCs with different mixtures of C6SH and C18SH at 303K, presented in Figure S8.*

`

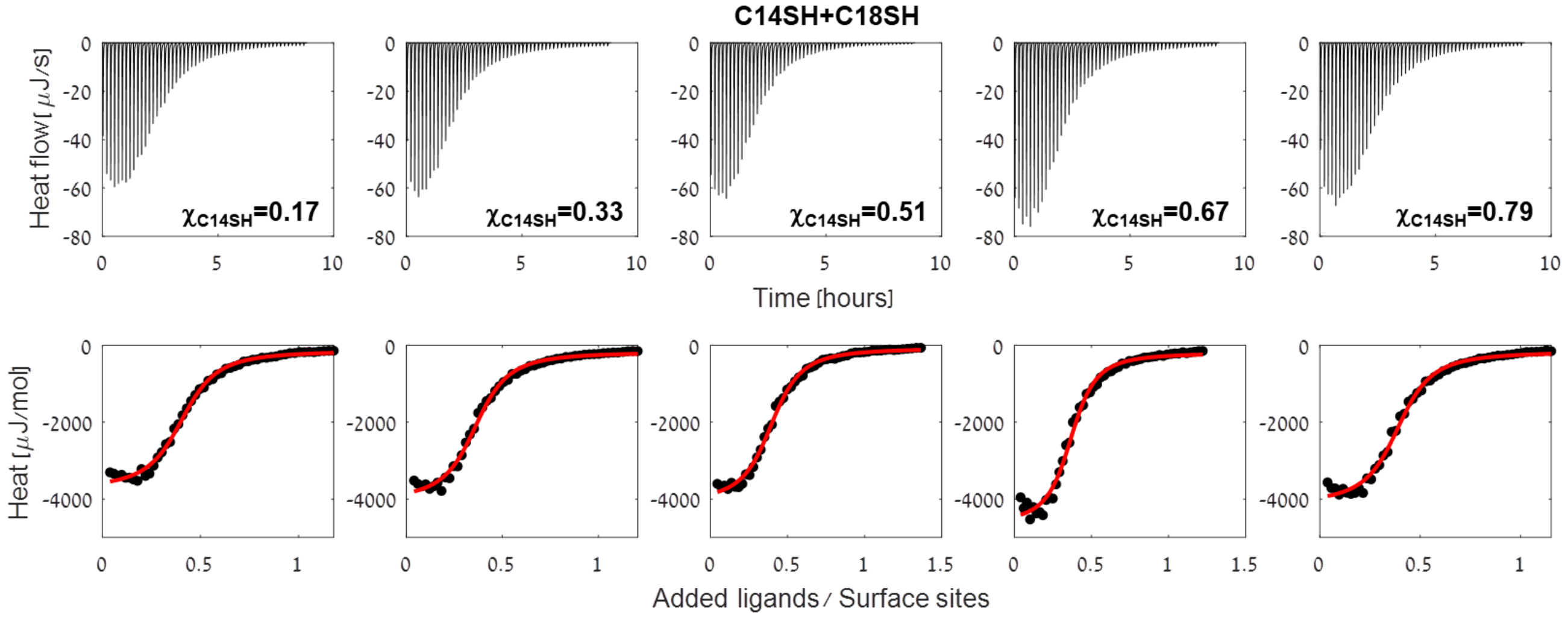

***Figure S9.*** *Real-time thermograms and the corresponding titration curves with their model fittings (red) for the ligand exchange reaction of oleate-coated CdSe NCs with different mixtures of C14SH and C18SH at 303K.*

| **χ(C14SH)** | **Ligand [mM]** | **Surface sites [mM]** | **ΔH [kJ/mol]** | **ΔS [J/molK]** | **ΔG [kJ/mol]** | **N** |
|---|---|---|---|---|---|---|
| **0.17** | 25 | 6.5 | -29.0 | -71 | -7.5 | 0.44 |
| **0.33** | 23 | 5.8 | -31.1 | -79 | -7.2 | 0.39 |
| **0.51** | 22 | 5.0 | -31.8 | -81 | -7.2 | 0.44 |
| **0.67** | 23 | 5.7 | -36.0 | -94 | -7.6 | 0.39 |
| **0.79** | 23 | 6.1 | -32.0 | -80 | -7.6 | 0.42 |

***Table S8.*** *A table summarizing the thermodynamic parameters extracted from the single-site model fit of the titration curves for the ligand exchange reaction of oleate-coated CdSe NCs with different mixtures of C14SH and C18SH at 303K, presented in Figure S9.*

`

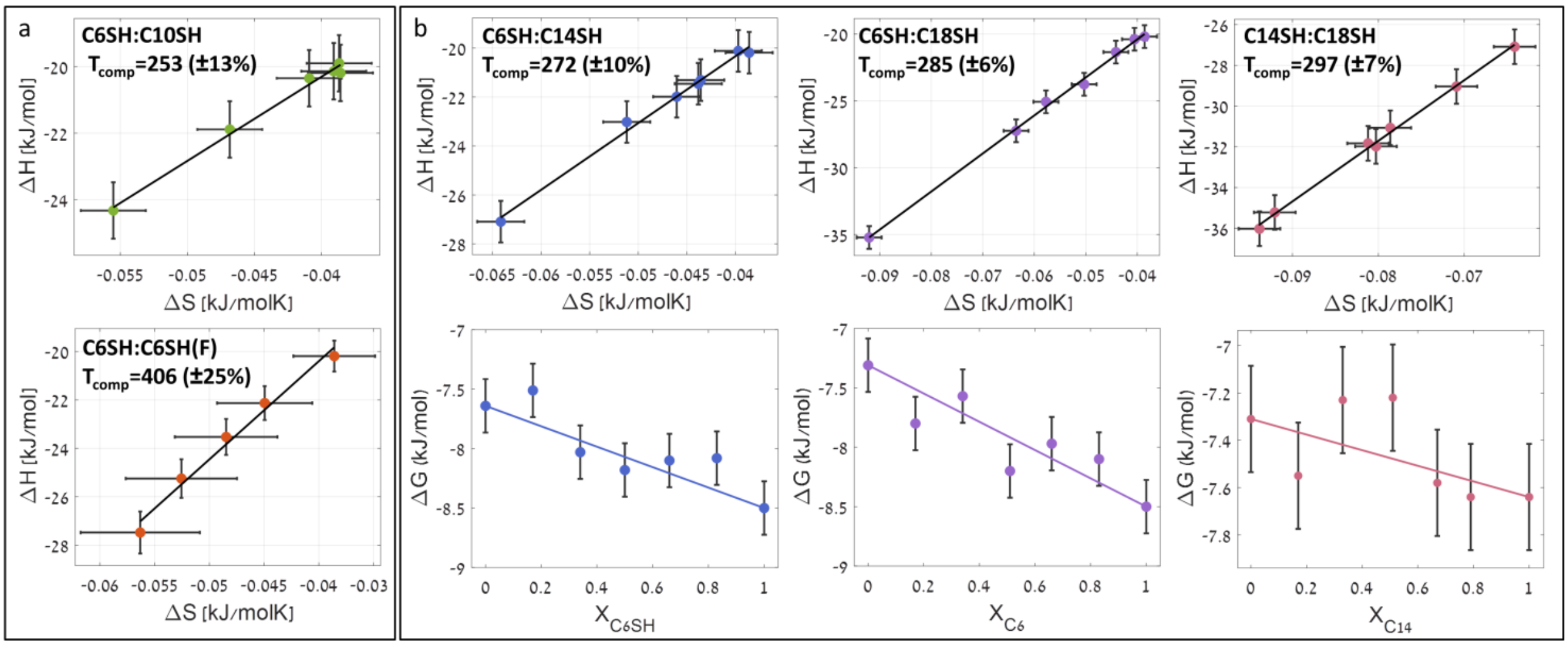

***Figure S10.*** *(a) Enthalpy-entropy compensation (EEC) plot for C6SH:C10SH (top) and C6SH:C6SH(F) (bottom) binary compositions, according to the ITC results presented in Table S4 and Table S5, respectively). The compensation temperature ($T_{comp}$) in the C6SH:C6SH(F) system deviates from the experimental temperature (303K), correlating with the changes in the measured ΔG values which indicate on poor compensation (Figure 1 in the main text). (b) EEC plot (top) and ΔG experimental values (dots) vs. the calculated LC (line), for the investigated C6SH:C14SH (blue), C6SH:C18SH (purple), C14SH:C18SH (pink) binary systems. The observed EEC is correlated with the minor changes observed in ΔG. The ITC-extracted values of ΔG correspond with the LC values, considering the errors.*

`

## 5. Additional surface characterization

Thermogravimetric analysis (TGA) was used to quantify the changes in the organic coverage of the NCs upon ligand exchange, where the differences in the overall mass loss and in the shape of the TGA thermograms indicate the changes in the surface ligand layer composition. Representative TGA results for the NCs with their native oleate ligands and upon ligand exchange with pure C6SH, pure C10SH as well as with $C6SH_{0.5}$:$C10SH_{0.5}$ and $C6SH_{0.66}$:$C10SH_{0.34}$ binary compositions are presented in Figure S11. The initial oleate coverage of the NCs was determined via the mass loss up to 500$^{o}$C, which is attributed to any organic species present in the sample. According to the thermogram, 44% of the total mass was organic, consistent with a full surface coverage and a 1:1 binding ratio (all Cd surface sites are bound to a single oleate ligand).

The post-ITC samples were analyzed similar to our previous report.[2] Prior to the analysis, the post-ITC samples were purified from excess free ligands by multiple cycles of precipitation and re-dispersion process using toluene (solvent) and ethanol (anti-solvent). As mentioned elsewhere,[2] the ITC conditions allow only partial ligand exchange, due to the small amounts of alkylthiol ligands added to the NCs in each titration point, resulting in an overall excess of alkylthiols to oleate of 1.5-2, which is insufficient to induce complete ligand exchange. This stands in contrast to the conditions for full ligand exchange which require an immediate disturbance to the system by the quick addition of a large excess of the exchanging ligand. Hence, the final surface coverage is slightly more complex, as more than one ligand type is involved. As can be observed from Figure S11, for the oleate coated NCs (red), the main mass loss is above 310$^{o}$C, while the mass loss of the alkylthiolate in the post-ITC samples is mostly lower than that. Thus, similar to our previous report, the thermograms of the post-ITC samples were divided into 2 regions: 110-315$^{o}$C, which is attributed to bound alkylthiolate (Figure S11, blue region), and 315-500$^{o}$C, which is attributed to bound oleate (Figure S11, red region). The C6SH:C10SH ligand ratio in the binary shell samples was taken as the ratio in added ligand solution. Additional analysis for the binary shell composition is discussed in details in the next paragraph. The ligand composition results for all post-ITC samples are summarized in Table S9. According to the calculated fraction of each ligand, 82%±2 of the total ligand coverage is alkylthiolate and 18%±2 is oleate. We note that all TGA analysis can have up to 10% of inaccuracy due to the different purification efficiency of each sample. Considering this error and the ligand composition calculation, we conclude that exchange with ligand mixtures sparsely affect the final shell coverage and the exchange ratio. Moreover, since our

`

thermodynamic analysis focuses on comparison between different ITC measured samples (including the comparison of the ligand exchange with binary mixtures with the linear combination of the pure ligands parameters), we can discard the thermodynamic effect of oleate remains on the ligand shell, as it present in all samples with similar fractions.

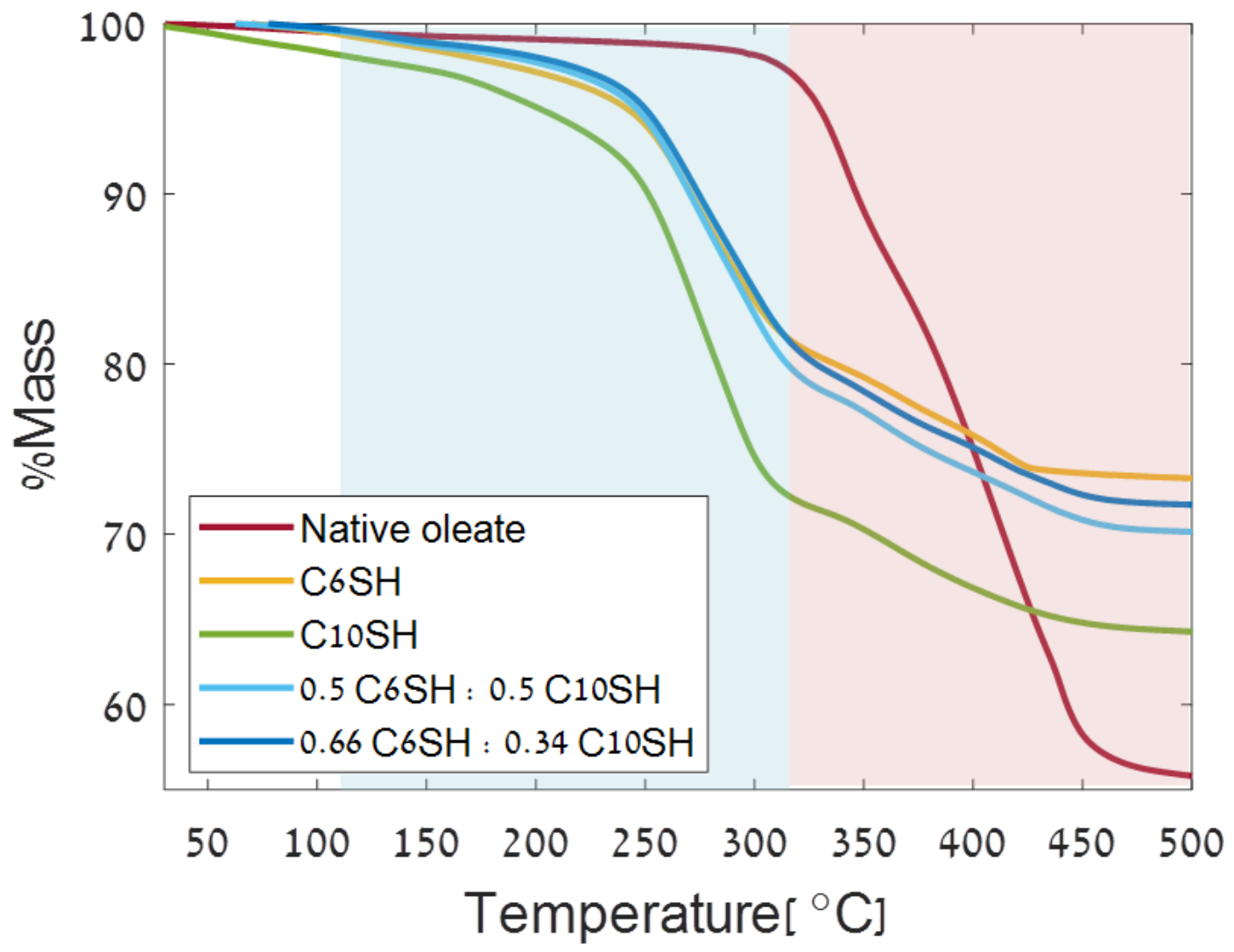

***Figure S11.*** *TGA thermograms of purified CdSe NCs before (red) and after ITC-preformed ligand exchange with pure C6SH (yellow), pure C10SH (green) and* $C6SH_{0.5}$*:*$C10SH_{0.5}$ *(light blue) and* $C6SH_{0.66}$*:*$C10SH_{0.34}$ *(dark blue) mixtures. The differences in the temperature of inflection points and the loss of organic mass, indicate the changes in the surface ligands.*

| Surface ligand | %Thiolate (110-315 ºC) | %Oleate (315-500 ºC) | Total coverage |
|---|---|---|---|
| **C6SH** | 80% | 16% | 96% |
| **C10SH** | 90% | 17% | 107% |
| **0.5 C6SH : 0.5 C10SH** | 74% | 19% | 94% |
| **0.66 C6SH : 0.34 C10SH** | 72% | 19% | 91% |

***Table S9.*** *Post-ITC ligand composition, as calculated from the TGA results presented in Figure S11.*

As described in the main text, the ITC data analysis in this work assumes that the ligand ratio on the NCs was as in the titrant added to the solution. To estimate the ligand ratio on the NC surface, we used the Langmuir model for competitive adsorption.[6] According to this model, the ratio of the surface absorbed molecules can be derived using ΔG values extracted from the experimental ITC curves. Specifically, the surface coverage of each component is given by:

. According to this model, we can estimate the ratio of the surface absorbed molecules, based on the ΔG values extracted from the experimental ITC curves. In this model. The surface coverage of each component is given by:

$$\text{(S25)} \quad \theta_A = \frac{K_{eq,A}X_A^{sol}}{1+K_{eq,A}X_A^{sol}+K_{eq,B}X_B^{sol}} \; ; \; \theta_B = \frac{K_{eq,B}X_B^{sol}}{1+K_{eq,A}X_A^{sol}+K_{eq,B}X_B^{sol}}$$

where $K_{eq,i}$ is the equilibrium constant for the exchange with pure ligand i (A or B, extracted from the ITC measurements with one ligand) and $X_i^{sol}$ is the molar fraction of ligand i, added to the solution. Based on eq. (S25) we calculated the molar fraction of each ligand on the NC surface. Results for C6SH:C10SH binary compositions are shown in Figure S12, as a representative system. The observed deviation between the molar fraction added to the solution (green) and that estimated on the NC surface (grey) is minor. Hence, we have validated using the titrant's molar fraction in the numerical model used for extracting the interaction parameters between the ligands (mentioned in the main text and in the next section).

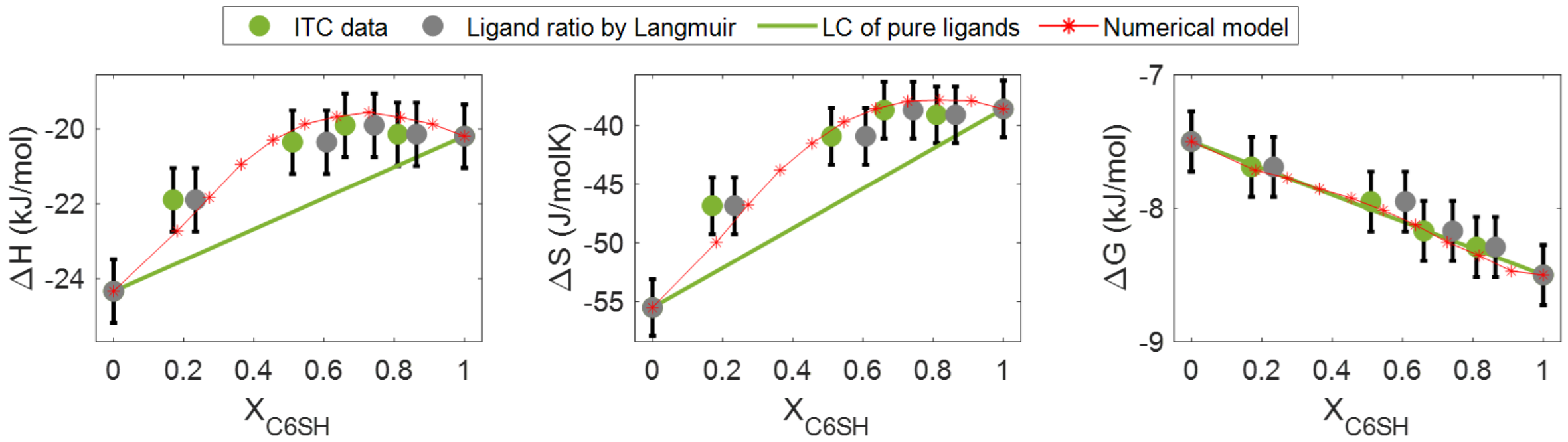

***Figure S12.*** *ITC extracted thermodynamic parameters: enthalpy, entropy, and Gibbs free energy, for the exchange with C6SH (XC6SH=1), C10SH (XC6SH=0) and their binary composition (0<XC6SH<1). Data are presented as a function of C6SH molar fraction in solution (green) and on NC surface based on a competitive Langmuir model calculation (grey). Solid green line represents the calculated linear combination (LC) of the pure ligands. Red asterisk and line represent the numerical model.*

`

# 6. Models of binary ligand shell

## 6.1 Ideal mixture model

We define an ideal mixture as one where there is no excess enthalpy upon mixing ($\Delta H_{mix}=0$), and the mixing process is spontaneous due only to the increase in the configurational entropy. The configurational entropy is based on the enumeration of all the available states for spatially organizing the mixture. For an ideal binary mixture on a lattice with total $L_A+L_B$ sites ($L_A$ and $L_B$ are the numbers of each ligand), the ideal configurational entropy of mixing is:

$$\text{(S26)} \quad \Delta S_{mix}^{conf,id} = R\ln\frac{(L_A+L_B)!}{L_A!L_B!} \approx -R(L_A+L_B)\left(X_{L_A}\ln X_{L_A} + X_{L_B}\ln X_{L_B}\right)$$

where $X_{LA}$ and $X_{LB}$ are the molar fractions of the two components.

## 6.2 Regular mixture model

The regular mixture model is the simplest mean field model that allows to describe non-ideal mixing between two or more components. The model assumes an excess enthalpy and entropy of mixing ($\Delta H_{mix}$ and $\Delta S_{mix}$, respectively) that are determined by the averaged environment of the components in the mixture. The excess enthalpy is modeled using an interaction parameter $\xi_H$ that indicates the invested or released energy involved with pairing two nearest neighbors of different ligands from pairs of similar ligands:

$$\text{(S27)} \quad \Delta H_{mix} = \xi_H X_{L_A} X_{L_B}$$

The excess entropy is determined by a configurational entropy that is approximated as the ideal contribution, eq.(S26), and a non-configurational entropy that includes an entropic interaction parameter $\xi_s$ that indicates the corresponding change in intra and inter-ligand degrees of freedom upon ligands pairing:

$$\text{(S28)} \quad \Delta S_{mix}^{nc} = \xi_S X_{L_A} X_{L_B}$$

The total free energy of mixing $\Delta G_{mix}$ is the sum of excess enthalpy and entropy (eqs. (S26)-(S28)):

$$\text{(S29)} \quad \Delta G_{mix} = \Delta H_{mix} - T\left(\Delta S_{mix}^{conf,id} + \Delta S_{mix}^{nc}\right)$$

`

6.3 Resolving thermodynamic parameters using thermodynamic integration

Applying the regular solution mean field model resulted in an inconsistency in the expected ligand shell structures: while the similar chemical nature of the binary ligands (C6SH and C10SH) suggests some degree of mixing, the model-calculated Gibbs free energy values suggested phase separation. In order to resolve this discrepancy, we fit our data to numerically exact solutions of the model as applied to a small system.

The free energy of a binary system is calculated considering the interactions between all nearest-neighboring ligand pairs $L_iL_j$, according to the Hamiltonian:

(S30) $\mathcal{H} = \sum_{i,j}\left(\chi_{G,i,j} L_i L_j\right)$

where the sum is performed over all nearest neighbors and the inter-ligand interaction parameter, $\chi_G$, includes enthalpic and entropic terms,

(S31) $\chi_G = \chi_H - T\chi_S$

Using the thermodynamic integration methodology,[7] we calculate $\chi_G$, which provides excess free energy for mixing according to:

(S32) $\Delta G_{mix,ex} = \int_0^1 \langle \frac{\partial \mathcal{G}(\lambda)}{\partial \lambda} \rangle_\lambda \, d\lambda$

where $\lambda$, the integration parameter, represents the path for changing the Hamiltonian from a reference system ($\lambda$=0) to the system of interest ($\lambda$=1), and $\mathcal{G}$ is the system's free energy, calculated as the thermal average of $\mathcal{H}$. Note that here our reference point is chosen to be the ideally mixed state, for which the configurational mixing entropy is described by eq.(S26):

(S33) $\Delta G_{mix}^{id} = -RTln\frac{(L_A+L_B)!}{L_A! L_B!}$

where T is the experimental temperature (303K). Hence, the total $\Delta G_{mix}$ corresponding to the experimentally measured value $\Delta G_{mix}$=0 is given by summing eqs. (S32) and (S33).

To simulate the NC surface, a square 11x11 grid (121 sites) with periodic boundary conditions was used to fit the total number of experimentally known surface sites (127, similar results were achieved for 12x12 grid). For a given number of $L_A$ and $L_B$, the system was first initialized to a randomly mixed state (similar results were achieved when starting with a phase separated state). Then, the system was allowed to reach equilibrium by performing a large number of Monte Carlo steps. According to the Metropolis algorithm,[8] at each step, a trial swap between two ligands was suggested and the change in the system energy upon switching, $\Delta U$, was calculated whereby each ligand interacts with its 4 neighbors: similar ligand pairs interact with 0 energy ($\chi_{G,AA}$=$\chi_{G,AB}$=0), and non-similar ligands interact with $\lambda\chi_G$. Swapping was allowed if it resulted in a decrease in the free energy, weighted by the Boltzmann probability,

`

(S34) $P_{switch} \propto \exp(\Delta\mathcal{G}\ /RT\ )$

After multiple steps, equilibrium was achieved, and the total energy of the system was calculated to give $\mathcal{G}(\lambda)$ for a specific λ. The procedure was repeated multiple times to give an averaged $\mathcal{G}(\lambda)$, which was later used for calculating $\Delta G_{mix}$ as described above.

The simulation was performed for several ratios of $L_A$ and $L_B$, to give $\Delta G_{mix}$ over the full range of ligand molar fractions. As detailed in the main text, the interaction parameter $\chi_G$ which correspond to $\Delta G_{mix}$=0 is 3.4RT, and this interaction parameter was used for simulating the ligand shell structure at equilibrium, using the Monte Carlo algorithm described here. The number of non-similar pairs (Figure 2 in the main text) and the average cluster area shown in Figure S13 were extracted from the simulated grid at equilibrium.

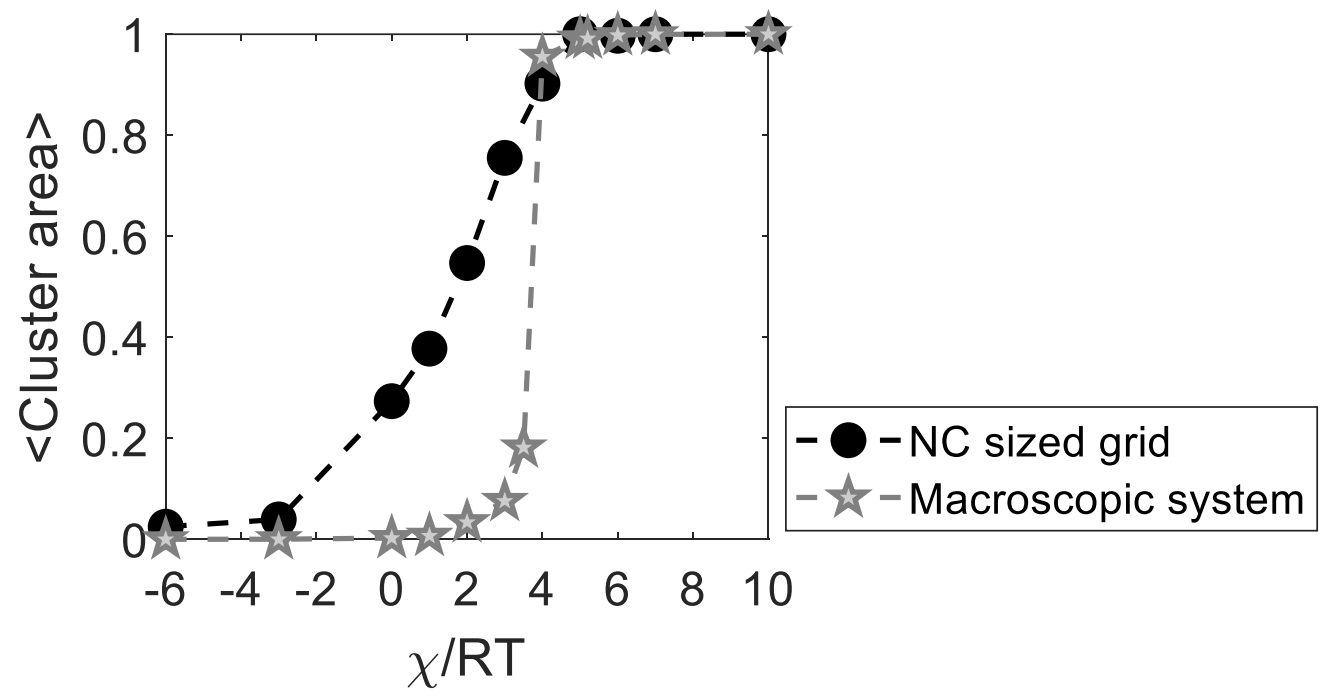

***Figure S13.*** *Average cluster area as a function of the total interaction parameter, $\chi_G$ in simulation. Results are normalized to the maximal cluster size in the phase separation state, RT is the thermal energy, and the molar fraction of the surface ligands shown here is X=0.5. Results are presented for the NC sized grid (11x11), which exhibit a gradual change, and for the macroscopic system approaching the thermodynamic limit (111x111).Cluster area changes gradually for the NC, but increases abruptly around the reported phase transition point (~3.5RT)*[9,10] *in the large system.*

Fits for the enthalpy and the entropy (for each binary ligand set) are derived from the resulting $\chi_G$. To resolve the entropic contribution to the free energy, we used the same thermodynamic integration methodology described above, but this time $\Delta G_{mix}$ was calculated for a temperature range of 297K to 309K (around the experimental temperature of 303K). The temperature dependence of $\chi_G$ was considered according to eq. (S31), allowing to determine $\chi_H$ and $\chi_S$ that best reproduce the experimental results. The enthalpy and the entropy for mixing were calculated from the temperature dependent simulation according to the van't Hoff relation:

(S35) $\Delta S_{mix} = -\frac{dG_{mix}}{dT};\ \Delta H_{mix} = \Delta G_{mix} + T\Delta S_{mix}$

`

As described in the main text, $\Delta S_{mix}$ is composed of configurational and non-configurational terms. $\Delta S_{mix}^{nc}$ is extracted by calculating the interactions (derived from $\chi_S$) between all $L_AL_B$ pairs at equilibrium, for all simulated $X_{C6SH}$. Then, $\Delta S_{mix}^{conf}$, is extracted by subtracting $\Delta S_{mix}^{nc}$ from the total $\Delta S_{mix}$. Since $\Delta S_{mix}^{conf}$ represents the entropy of the ligand organization, which is directly derived from $\chi_G$, systems with similar $\chi_G$ exhibit similar $\Delta S_{mix}^{conf}$.

For the C6SH:C6SH(F) mixture, a similar procedure was applied with the requirement of positive total $\Delta G_{mix}$ so as to match the ITC results.

For the C14SH:C18SH mixture, the simulation included in addition a linear dependence of $\chi_H$ and $\chi_S$ on ligand's molar fractions, as detailed in the main text. For completeness, we applied similar simulations also for the other investigated binary mixtures of C6SH:C10H, C6SH:C14SH, and C6SH:C18SH (Figure S14), which exhibit deviations from the fitting provided by constant $\chi_H$ and $\chi_S$. However, we note that since the deviations from the original model were minor, the new model should be analyzed with caution so as to avoid over-interpretation. Comparing the fitting-extracted $\omega$ and $\eta$ (Figure S14g and S14h, respectively), it is noticeable that the long ligands at each binary system induce loss of interaction energy and gain in entropy upon mixing. The parameters of C14SH, which represent the long ligand in the C6SH:C14SH system and the short ligand in the C14:SH:C18SH system, switch sign between both system, as the C14SH gain (loss) interactions and loss (gain) entropy upon mixing with the longer C18SH (shorter C6SH). The parameters for the C10SH and C6SH ligands are low and may vary due to using multiply fitting parameters to a system that could already be reasonably fitted with a single one (constant $\chi$).

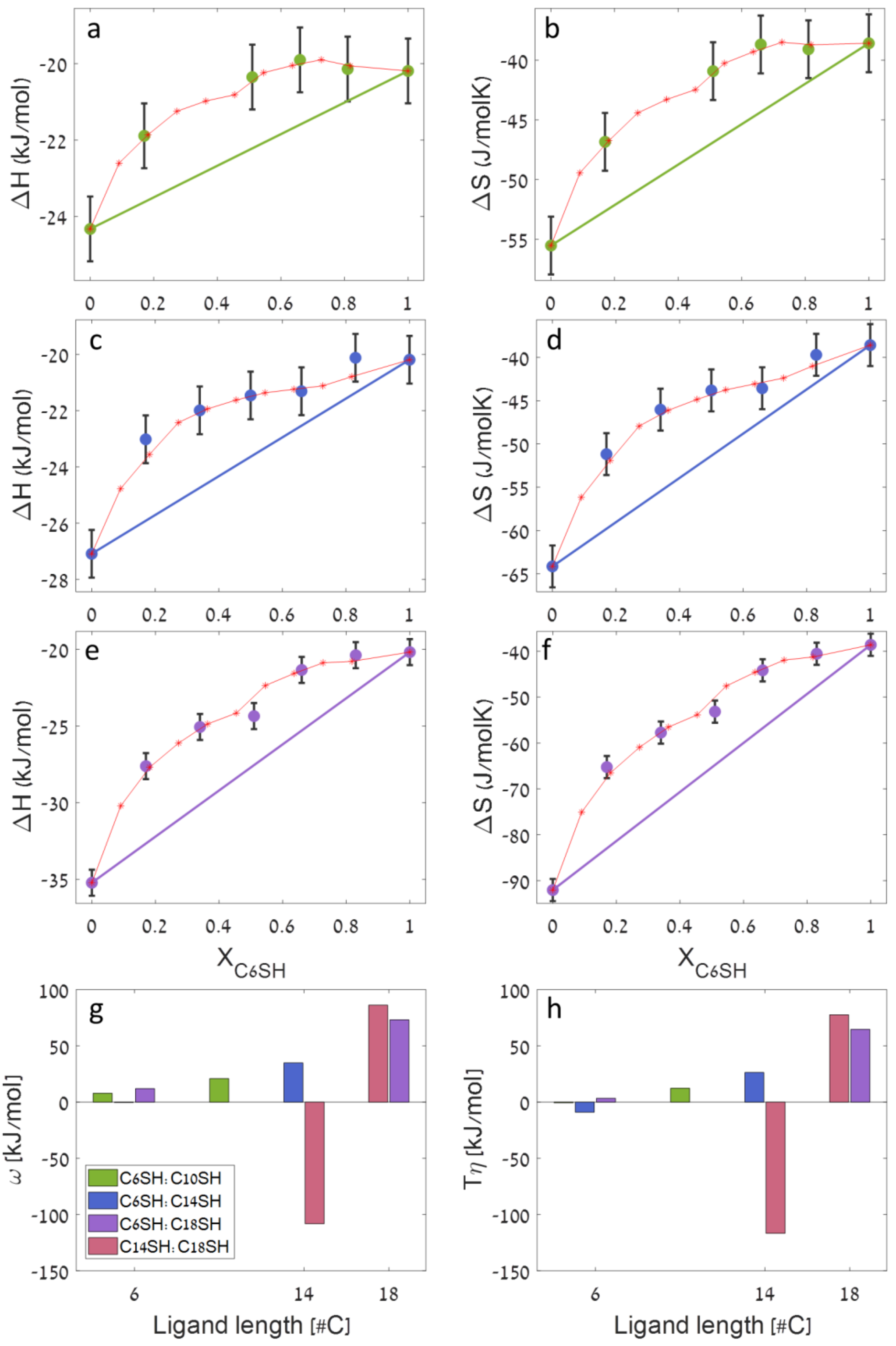

***Figure S14.*** *ITC results (dots) and numerical fitting (red asterisks and lines) considering a composition dependent interaction parameters for (a-b) C6SH:C10SH, (c-d) C6SH:C14SH, and (e-f) C6SH:C18SH binary compositions. (g) Fitted enthalpic and (h) entropic interaction parameter coefficients for each ligand in the binary compositions.*

`